\documentclass[final]{l4dc2024}

\usepackage{xcolor} 

\usepackage{bm}
\usepackage{amsmath}
\usepackage{multicol}
\usepackage{graphicx}
\usepackage{multirow}

\newcommand{\R}{\mathbb{R}}
\newcommand{\X}{\mathcal{X}}
\newcommand{\T}{\mathcal{T}}
\newcommand{\x}{\bm{x}}
\renewcommand{\u}{\bm{u}}

\newcommand{\A}{\mathop{\int\!\!\!\int}\limits_{A}}

\usepackage{algorithm} 
\usepackage{algpseudocode} 
\usepackage{tikz}
\definecolor{ForestGreen}{rgb}{0.0, 0.5, 0.0}
\newcommand{\redx}{%
\color{red}
\tikz[scale=.7] {
    \draw[line width=0.7,line cap=round] (0,0) to [bend left=6] (1,1);
    \draw[line width=0.7,line cap=round] (0.2,0.95) to [bend right=3] (0.8,0.05);
}}
\newcommand{\greencheckmark}{
\color{ForestGreen}
\tikz[scale=.7] {
    \draw[line width=0.7,line cap=round] (0.25,0) to [bend left=10] (1,1);
    \draw[line width=0.8,line cap=round] (0,0.35) to [bend right=1] (0.23,0);
}}

\usepackage{float}

\allowdisplaybreaks
\title[PDE Control Gym]{PDE Control Gym: A Benchmark for Data-Driven Boundary Control of Partial Differential Equations}
\usepackage{times}

\author{%
 \Name{Luke Bhan}\thanks{equal contribution} \Email{lbhan@ucsd.edu}\\
 \addr University of California, San Diego \\  
  \Name{Yuexin Bian}\footnotemark[1] \Email{yubian@ucsd.edu}\\
 \addr University of California, San Diego  \\ 
  \Name{Miroslav Krstic} \Email{krstic@ucsd.edu}\\
 \addr University of California, San Diego \\
 \Name{Yuanyuan Shi} \Email{yyshi@ucsd.edu} \\ 
 \addr University of California, San Diego 
}

\begin{document}

\maketitle

\begin{abstract}%
Over the last decade, data-driven methods have surged in popularity, emerging as valuable tools for control theory. As such, neural network approximations of control feedback laws, system dynamics, and even Lyapunov functions have attracted growing attention. 
With the ascent of learning based control, the need for accurate, fast, and easy-to-use benchmarks has increased. In this work, we present the first learning-based environment for boundary control of PDEs.   
In our benchmark, we introduce three foundational PDE problems - a 1D transport PDE, a 1D reaction-diffusion PDE, and a 2D Navier–Stokes PDE - whose solvers are bundled in an user-friendly reinforcement learning gym. With this gym, we then present the first set of model-free, reinforcement learning algorithms for solving this series of benchmark problems, achieving stability, although at a higher cost compared to model-based PDE backstepping. With the set of benchmark environments and detailed examples, this work significantly lowers the barrier to entry for learning-based PDE control - a topic largely unexplored by the data-driven control community. The entire benchmark is available on \href{https://github.com/lukebhan/PDEControlGym}{Github} along with detailed \href{https://pdecontrolgym.readthedocs.io/en/latest/}{documentation} and the presented reinforcement learning models are \href{https://huggingface.co/lukebhan/PDEControlGymModels}{open sourced}.
\end{abstract}

\section{Introduction}
As learning-based control has exploded across both academia and industry, the need for fast, and accurate bench-marking is heightened. For example, perhaps the most visible impact of proper bench-marking is in the field of computer-vision resulting in 15-years of breakthrough results from AlexNet \cite{NIPS2012_c399862d} to neural radiance fields (NeRFs) \cite{mildenhall2020nerf}. Despite this, the control community has, justifiably, forgone consistent efforts in bench-marking as the community spawned from an applied mathematics perspective where the focus was behind \textit{provable} stability guarantees. However, given the recent exploration surrounding data-driven control methods \cite{9903316, 10336939}, designing fast, well-documented, and challenging benchmarks is of utmost importance to ensure new learning-based control approaches are consistently advancing the state of the art. 

In this work, we develop a benchmarking suite for learning-based boundary control of PDEs. Boundary control of PDEs is of elevated importance compared to control across the full domain as many real world problems \textit{cannot} control the PDE across the entire domain, but only at the boundary input. Thus, boundary control is physically more realistic as the actuation and sensing are generally non-intrusive \cite{doi:10.1137/1.9780898718607}. For example, in fluid flows, the engineer only gets control access to the surrounding walls containing the fluid  \cite{6315422} or in temperature manufacturing~\cite{bian2024ventilation}, the engineer is typically unable to set the temperature of entire plate, but only a specific edge. Furthermore, boundary control is extremely powerful in modeling macro-level traffic congestion \cite{trafficHuanYu} as modern highways typically only enable actuation and sensing at the on/off ramps.
Lastly, we briefly mention that boundary control is one of the only approaches to handle chemical and combustion processes \cite{IZADI201541} and moving boundary problems such as the Stefan Problem with applications to both 3D printing \cite{koga} and the excitation of neuron growth for abating neurological diseases such as Parkinson's and Alzheimer's \cite{DEMIR2024111669}. 
However, for a majority of these applications, each researcher typically develops their own simulations and thus there is no standard library with a universal set of problems to test new algorithms. Furthermore, for even the standard model-based algorithms such as PDE backstepping, it is challenging to implement the control algorithm as this typically requires the solution of a \textit{separate} Goursat PDE \cite{10384080}. Thus, in this work, we introduce the first library containing a set of general PDE control problems and implementations of their corresponding model-based control algorithms that can be easily modified to fit the wide array of aforementioned target applications.  
   

\paragraph{Contributions} This paper has three main contributions. First, we introduce, design, and formalize the first benchmark suite for PDE control including 3 classical problems ranging from boundary stabilization for 1D transport (hyperbolic) and reaction-diffusion (parabolic) PDEs to trajectory following for the 2D Navier-Stokes PDEs. Along with the proposal of the PDE control benchmarking suite, we parameterize the numerical scheme implementations as RL gyms - effectively decoupling the PDE solvers from the controller design, enabling the use of \emph{any pre-implemented learning algorithm} for PDE control. Second, utilizing our benchmark suite, we train the first set of \emph{model-free} RL controllers which effectively stabilize hyperbolic and parabolic PDE problems and achieve effective tracking for the 2D Navier-Stokes equations. We then compare the resulting controllers to classical algorithms such as PDE backstepping and adjoint-based optimization highlighting the trade-offs between performance. Lastly, we provide extensive documentation and numerous examples for the training of RL controllers, implementation of classical control algorithms and of course the integration of new PDE control problems into the benchmark suite.



\section{Related Work}\label{section:relatedWork}
\subsection{Learning-based PDE benchmarks}
\begin{table}[ht]
\centering
\resizebox{\textwidth}{!}{%
\begin{tabular}{|l|c|c|c|c|c|c|}
\hline
\textbf{\begin{tabular}[c]{@{}l@{}}Benchmarking for machine\\ learning in PDEs\end{tabular}} & \multicolumn{1}{l|}{\textbf{\begin{tabular}[c]{@{}l@{}}Compilation of a\\ premade dataset\end{tabular}}} & \multicolumn{1}{l|}{\textbf{Supports control}} & \multicolumn{1}{l|}{\textbf{\begin{tabular}[c]{@{}l@{}}Differentiable PDE\\ solver\end{tabular}}} & \multicolumn{1}{l|}{\textbf{\begin{tabular}[c]{@{}l@{}}Supports custom\\ PDEs\end{tabular}}} & \multicolumn{1}{l|}{\textbf{\begin{tabular}[c]{@{}l@{}}Supports reinforcement\\ learning\end{tabular}}} & \multicolumn{1}{l|}{\textbf{\begin{tabular}[c]{@{}l@{}}Implementation of\\ model-based control\end{tabular}}} \\ \hline
PhiFlow \cite{Holl2020Learning}                                                                                     & \begin{tabular}{l} \redx \end{tabular}                                                                                     & \begin{tabular}{l} \greencheckmark \end{tabular}                 & \begin{tabular}{l} \greencheckmark \end{tabular}                                                                    & \begin{tabular}{l} \greencheckmark \end{tabular}                                                               & \begin{tabular}{l} \redx \end{tabular}                                                                                    & \begin{tabular}{l} \redx \end{tabular}                                                                                          \\ \hline
\begin{tabular}[c]{@{}l@{}}PDEBench \cite{NEURIPS2022_0a974713} \\ (created from PhiFlow)\end{tabular}                   & \begin{tabular}{l}\greencheckmark \end{tabular}                                                                          & \begin{tabular}{l}\redx\end{tabular}                           & \begin{tabular}{l} \redx \end{tabular}                                                                              & \begin{tabular}{l} \redx \end{tabular}                                                                         & \begin{tabular}{l} \redx \end{tabular}                                                                                    & \begin{tabular}{l} \redx \end{tabular}                                                                                          \\ \hline
\begin{tabular}[c]{@{}l@{}}PDEArena \cite{gupta2022towards}\\  (created from PhiFlow)\end{tabular}                   & \begin{tabular}{l} \greencheckmark \end{tabular}                                                                           & \begin{tabular}{l} \redx \end{tabular}                           & \begin{tabular}{l} \redx \end{tabular}                                                                              & \begin{tabular}{l} \redx \end{tabular}                                                                         & \begin{tabular}{l} \redx \end{tabular}                                                                                    & \begin{tabular}{l} \redx \end{tabular} \\ \hline

PDE Control Gym (Ours)                                                                          & \begin{tabular}{l} \redx \end{tabular}                                                                                     & \begin{tabular}{l} \greencheckmark \end{tabular}                 & \begin{tabular}{l} \redx \end{tabular}                                                                              & \begin{tabular}{l} \greencheckmark \end{tabular}                                                               & \begin{tabular}{l} \greencheckmark \end{tabular}                                                                          & \begin{tabular}{l} \greencheckmark \end{tabular}                                                                                \\ \hline
\end{tabular}%
}
\caption{Comparison of benchmarks for machine learning in PDEs.}
\end{table}
Currently, to the author's knowledge, there are no benchmarking suites focused on \emph{boundary control} of PDEs as most benchmarks such as \cite{NEURIPS2022_0a974713, gupta2022towards} present datasets for learning PDE solution maps from initial conditions. These benchmarks are typically used for comparing neural network-based PDE solvers like neural operators \cite{lu2021, li2021fourier} and PINNs \cite{RAISSI2019686}. Although these benchmarks are effective, they do not allow users to incorporate boundary control or alter the PDEs within their datasets. The $\phi_{\text{Flow}}$ framework \cite{Holl2020Learning} provides a flexible PDE solver that supports automatic differentiation for the calculation PDE derivatives to be used in control algorithms. While adaptable for PDE control, it does not natively support RL algorithms and lacks a set of model-based controllers for learning-based control comparisons. Lastly, it is worth noting that while there are ODE control suites like \cite{duan2016benchmarking,DBLP:journals/corr/abs-1801-00690} tailored for ODE control tasks. Lastly, we mention a concurrent work that also integrates \textbf{in-domain} PDE control into a RL library \cite{zhang2024controlgym}; however, the class of problems they focus on - namely in-domain control - is different from the boundary control problems we benchmark in this gym. Thus, the PDE control gym, to our knowledge, represents the \textit{first} PDE-focused benchmarking suite for learning-based boundary-control algorithms. 

\subsection{Learning as a tool for PDE control}
As with most scientific disciplines, machine learning has had a broad impact in PDE control. In 1D PDE problems, a series of work has been developed to use neural operators for approximating control feedback laws \cite{bhan2023neural,pmlr-v211-bhan23a, krstic2023neural, qi2023neural}, under a supervised learning framework with provable stability guarantees.  Furthermore, an optimal control approach using PINNs is explored in \cite{MOWLAVI2023111731}. Additionally, the closest paper to this work is by \cite{9568241} who presented the first exploration utilizing RL for PDE boundary control. However, they do not explore the benchmark PDE problems presented in this paper instead focusing on Aw-Rascale-Zhang (ARZ) traffic model. 

\subsection{Reinforcement learning in control}
Reinforcement learning (RL) has demonstrated significant success in various control applications, including robotics~\cite{brunke2022safe}, power systems~\cite{chen2022reinforcement}, and autonomous driving~\cite{kiran2021deep}. From a controls perspective, deep RL (DRL) algorithms learn feedback laws that maps observations (states) into actions (control inputs), typically via neural networks (NN). These RL controllers are trained to optimize specific reward functions, such as the $L^2$ spatial norm of states for stabilization tasks~\cite{9568241}.
The most appealing feature of deep RL is its \emph{model-free} nature, allowing it to control complex systems without requiring explicit model estimations. Consequently, RL has potential to outperform model-based control methods in highly complex tasks with hard-to-model dynamics. To demonstrate the use of PDE Control Gym, we conduct experiments using off-the-shelf RL algorithms implemented with Stable-Baselines3~\cite{raffin2021stablebaselines3}. We selected the off-policy soft actor-critic (SAC) \citep{haarnoja2018soft} and on-policy proximal policy optimization (PPO) \citep{schulman2017proximal} algorithms for their demonstrated efficiency in solving challenging continuous control tasks~\cite{duan2016benchmarking}.

\section{Formalization of PDE Control Problems}\label{sec:highlevelFormalize}
\subsection{General PDE control problem}
We consider a partial differential equation (PDE) defined on a domain $\X$, which can be either one-dimensional (1D), $\X = [0, 1] \subset \R$, or two-dimensional (2D), $\X = [0, 1] \times [0, 1] \subset \R^2$. The time domain is $\T = [0, T] \subset \R^+$.
Let $u(x, t), x \in \mathcal{X}, t \in \mathcal{T}$ describe the state of the system governed by the PDE according to the dynamics
\begin{equation}
        \frac{\partial u}{\partial t} = \mathcal{P}\left(u, \frac{\partial u}{\partial x}, \frac{\partial^2 u}{\partial x^2}, \ldots, U(t)\right)\,,
\end{equation}
where $\mathcal{P}$ is the partial differential equation(s) that model(s) the system dynamics, and $U(t)$ is the control function. Then, the goal of a PDE control problem is
 to optimize a cost function (e.g. regulate $u(x, t)$ to be the desired trajectory while reducing the control cost, stabilize the PDEs) from just boundary inputs. Note that in some methods such as PDE backstepping, optimization is forgone in favor of just asymptotic stabilization as the infinite dimensional nature of PDE control is extremely challenging.

\subsection{Markov decision processes (MDPs) for PDE control}
We give a brief overview of the components of the MDP governing both the 1D and 2D support problems. More details about the specific MDPs governing the examples in Section \ref{sec:experiments} can be found in the Appendix \ref{sec:hyperbolicExample} and \ref{sec:parabolicExample}.

\begin{table}[ht]
\centering
\resizebox{\textwidth}{!}{%
\begin{tabular}{l|cc|cc|cc}
                                                                            & \multicolumn{2}{l|}{\textbf{1D Hyperbolic}}                                                                                                                     & \multicolumn{2}{l|}{\textbf{1D Parabolic}}                                                                                                                      & \multicolumn{2}{l}{\textbf{2D Navier-Stokes}}                                                                 \\ \cline{2-7} 
\textbf{\begin{tabular}[c]{@{}l@{}}Supported \\ Configuration\end{tabular}} & \multicolumn{1}{l}{\textbf{Sensing ($o(t)$)}}                                  & \multicolumn{1}{l|}{\textbf{Actuation ($a(t)$)}}                               & \multicolumn{1}{l}{\textbf{Sensing ($o(t)$)}}                                  & \multicolumn{1}{l|}{\textbf{Actuation ($a(t)$)}}                               & \multicolumn{1}{l}{\textbf{Sensing ($o(t)$)}} & \multicolumn{1}{l}{\textbf{Actuation ($a(t)$)}} \\
\textbf{full-state}                                                         & {\color[HTML]{3166FF} \textbf{$u(x, t)$}}                                      & {\color[HTML]{3166FF} \textbf{$u(1, t)$}}                                      & {\color[HTML]{3166FF} \textbf{$u(x, t)$}}                                      & {\color[HTML]{3166FF} \textbf{$u(1, t)$}}                                      & {\color[HTML]{3166FF} \textbf{$u(x, y, t)$}}  & {\color[HTML]{3166FF} \textbf{$u(x, 1, t)$}}                  \\
\textbf{colloacted}                                                         & $u_x(1, t)$                                                                    & $u(1, t)$                                                                      & $u_x(1, t)$                                                                    & $u(1, t)$                                                                      & ---                                           & ---                                                           \\
\textbf{\begin{tabular}[c]{@{}l@{}}anti\\ collocated\end{tabular}}          & \begin{tabular}{l}$u(0, t)$\end{tabular}   & \begin{tabular}{l}$u(1, t)$\end{tabular}   & ---                                                                            & ---                                                                            & ---                                           & ---                                                           \\
\textbf{\begin{tabular}[c]{@{}l@{}}anti\\ collocated\end{tabular}}          & \begin{tabular}{l}$u_x(0, t)$\end{tabular} & \begin{tabular}{l}$u(1, t)$\end{tabular}   & \begin{tabular}{l}$u_x(0, t)$\end{tabular} & \begin{tabular}{l}$u(1, t)$\end{tabular}   & ---                                           & ---                                                           \\
\textbf{full-state}                                                         & $u(x, t)$                                                                      & $u_x(1, t)$                                                                    & $u(x, t)$                                                                      & $u_x(1, t)$                                                                    & $u(x, y, t)$                                  & $u(x, 0, t)$                                                  \\
\textbf{collocated}                                                         & $u(1, t)$                                                                      & $u_x(1, t)$                                                                    & $u(1, t)$                                                                      & $u_x(1, t)$                                                                    & ---                                           & ---                                                           \\
\textbf{\begin{tabular}[c]{@{}l@{}}anti\\ collocated\end{tabular}}          & \begin{tabular}{l}$u(0, t)$\end{tabular}   & \begin{tabular}{l}$u_x(1, t)$\end{tabular} & ---                                                                            & ---                                                                            & ---                                           & ---                                                           \\
\textbf{\begin{tabular}[c]{@{}l@{}}anti\\ collocated\end{tabular}}          & \begin{tabular}{l}$u_x(0, t)$\end{tabular} & \begin{tabular}{l}$u_x(1, t)$\end{tabular} & \begin{tabular}{l}$u_x(0, t)$\end{tabular} & \begin{tabular}{l}$u_x(1, t)$\end{tabular} & ---                                           & ---                                                           \\
\textbf{full-state}                                                         & ---                                                                            & ---                                                                            & ---                                                                            & ---                                                                            & $u(x, y, t)$                                  & $u(1, y, t)$                                                  \\
\textbf{full-state}                                                         & ---                                                                            & ---                                                                            & ---                                                                            & ---                                                                            & $u(x, y, t)$                                  & $u(0, y, t)$                                                 
\end{tabular}%
}
\label{table:problemConfigs}
\caption{Configurations for actuation and sensing supported by the PDE Control Gym for the three problems. Full state indicates the measurement is the entire PDE state, collated indicates that the sensing and measurement is done at the same boundary point, and anti-collocated indicates sensing and measurement are done at opposite boundary points. The configurations marked in \color[HTML]{3166FF}blue \color{black}  correspond to the experiment examples in Section \ref{sec:experiments}.}
\end{table}

\paragraph{State and Observation Space.} In all PDE control problems addressed, the state $s(t)$ at time $t$ is represented by the PDE value $u(x, t), x \in \mathcal{X}$. To enhance flexibility for different PDE tasks such as observer design, we have developed different partial state measurement settings, which we denote as the observation space $o(t)$. The types of sensing supported for each problem are detailed in Table \ref{table:problemConfigs}. Furthermore, we offer the ability to introduce custom noise functions (with built-in support for Gaussian noise). This allows users to simulate real-world sensor noise in their experiments.

\paragraph{Action Space.} The action $a(t) = U(t)$ for both the RL and control agents is determined by the actuation locations and boundary condition type. In 1D hyperbolic and parabolic systems, we consider both Neumann and Dirichlet boundary actuation $U(t) \in \mathbb{R}$ at either boundary, with four possible cases: 1) $u(0, t) = U(t)$, 2) $u(1, t) = U(t)$, 3) $u_x(0, t) = U(t)$, and 4) $u_x(1, t) = U(t)$. Considering the symmetry between the boundaries $x=0$ and $x=1$, there are eight distinct combinations for the 1D Hyperbolic PDE problem and six for the 1D Parabolic PDE problem, including an additional boundary condition at $u(0, t)$. These combinations are outlined in the first two sections of Table \ref{table:problemConfigs}. For the 2D Navier–Stokes problem, we consider Dirichlet-type boundary actuation on any of the four boundaries: top, bottom, left, and right. The gym also allows users to customize actuator positions, enabling research into optimizing both location and actuation type.

\paragraph{State Evolution.} To simulate the PDE system evolution with the control input $U(t)$, we use a first-order Taylor approximation for temporal evolution,
\begin{equation}
    u(t+\Delta t) = u(t) + \Delta t \cdot \mathcal{P}\left(u(t), \frac{\partial u}{\partial x}, \frac{\partial^2 u}{\partial x^2}, \ldots, U(t)\right)\,.
\end{equation}
Spatial derivatives are approximated using appropriate finite difference schemes and are explicitly given for each gym environment in the Appendix \ref{sec:hyperbolicFiniteDiff} and \ref{sec:parabolicFiniteDiff}. In practice, selecting the time step $\Delta t$ and spatial discretization $\Delta x$ for each problem requires careful consideration, particularly based on the number of approximated spatial derivatives. Nonetheless, we found that reasonable choices such as $\Delta x = 0.01$, $\Delta t=0.0001$ yield both fast and numerically stable results.

\paragraph{Reward.} 
Reward shaping plays a pivotal role in the training of RL algorithms. 
Generally speaking, for stabilization tasks, it is appropriate to employ a form of trajectory-based reward
\begin{eqnarray} \label{eq:reward1d}
	\int_0^{T} \left(\int_{x \in \mathcal{X}}\|u(x,t)\|^2 dx + \|U(t)\|^2\right) dt\,,
\end{eqnarray} which minimizes the state magnitudes and control efforts. 
For tracking tasks, a trajectory reward as
\begin{eqnarray} \label{eq:reward2d}
	\int_0^{T} \left(\int_{x \in \mathcal{X}}\|u({x},t)-u_{ref}({x}, t)\|^2 d {x}+ \|U(t)-U_{ref}(t)\|^2\right) dt\,,
\end{eqnarray} is a reasonable choice as it penalizes deviations from the reference trajectory given in both state $u_{ref}({x}, t)$ and control actions $U_{ref}(t)$. However, in practice, for the 1D hyperbolic and parabolic PDE problems, we found that the reward as given in \eqref{eq:reward1d} was insufficient for training and thus we use a specifically tuned reward that penalizes the difference of the $L_2$ norms between the current state and next state after action $a(t)$ (presented in Appendix \ref{sec:hyperbolicExample} and \ref{sec:parabolicExample}).

\section{Benchmark PDE Control Tasks}\label{sec:benchmarkProblems}
\subsection{1D Hyperbolic (transport) PDEs}\label{secsec:hyperbolic}
We consider the benchmark transport PDE in the form 
\begin{eqnarray} \label{eq:hyperbolic}
    u_t(x, t) &=& u_x(x, t) + \beta(x)u(0, t), 
\end{eqnarray}
for $x \in [0, 1), t \in [0, T]$.
Physically, \eqref{eq:hyperbolic} is a ``transport process (from $x=1$ towards $x=0$) with recirculation'' of the outlet variable $u(0,t)$. Recirculation causes \emph{instability} and the goal is to stabilize ``full-state'' recirculation from only boundary inputs. In practice, 
we consider the same PDE as studied in \cite{bhan2023neural} where $\beta$ is governed by the Chebyshev polynomial $\beta(x)=5\cos(\gamma \cos^{-1}(x))$ and Dirichlet actuation $u(1, t)=U(t)$. Classically, this PDE, with recirculation, has been a seminal-benchmark for PDE backstepping, as the 1D transport PDE can model a wide range of applications from chemical processes to shallow water waves and traffic flows \cite{KRSTIC2008750}.

\paragraph{Model-based backstepping control.}
The backstepping controller given by the following
\begin{eqnarray} \label{eq:bcskhyperbolic}
    U(t) &=& \int_0^1 k(1-y)u(y, t) dy\,,\\ 
    k(x) &=& -\beta(x) + \int_0^x \beta(x-y)k(y) dy \label{eq:kernel}\,,
\end{eqnarray}
for $x \in [0, 1]$.
\eqref{eq:bcskhyperbolic} results in stabilization of \eqref{eq:hyperbolic} \cite{KRSTIC2008750}. In practice, the backstepping kernel \eqref{eq:kernel} is implemented using the successive approximations approach although a Laplace transform approach is also viable \cite{doi:10.1137/1.9780898718607}. 

\subsection{1D Parabolic (reaction-diffusion) PDEs} \label{secsec:parabolic}
We consider the benchmark reaction-diffusion PDEs governed by recirculation function $\lambda(x)$ as 
\begin{eqnarray}
    \label{eq-PDE}
	u_t(x,t) &=& u_{xx}(x,t) + \lambda(x) u(x, t),
 \\  \label{eq-PDEBC0}
 u(0,t) &=& 0,
\end{eqnarray}
with Dirichlet or Neumann actuation at $x=1$. Again, instability is caused by the $\lambda(x)u(x, t)$ term otherwise the problem would simplify to the classical heat equation. This PDE appears in different applications ranging from a chemical tubular reactor \cite{9992759} to electro-chemical battery models \cite{moura2014adaptive} and diffusion in social networks \cite{wang2020modeling}.

\paragraph{Model-based backstepping control.} For the PDE \eqref{eq-PDE}, \eqref{eq-PDEBC0}, with Dirichlet boundary actuation $u(1, t)=U(t)$, the backstepping controller with full state measurement is given by the following \cite{1369395, Smyshlyaev2010},
\begin{equation}\label{eq-bkstfbkk}
    U(t) = \int_0^1 k(1,y) u(y,t) dy.
\end{equation}
where $k(x, y) \in C^2(\Tilde{\mathcal{T}})$, $\Tilde{\mathcal{T}} = \{0 \leq y \leq x \leq 1\}$.
\begin{eqnarray}
\label{eq-kPDE}
k_{xx}(x,y) - k_{yy}(x,y) &=& \lambda(y) k(x,y), \quad \forall (x,y)\in \breve{\mathcal T}\,,
\\ \label{eq-kPDEBC0}
k(x,0) &=& 0\,,
\\ \label{eq-kPDEBCx}
k(x,x) &=& -{\frac{1}{2}} \int_0^x \lambda(y) dy\,,
\end{eqnarray}
where $\breve{\mathcal T} = \{0< y\leq x <1\}$.

\subsection{2D Navier-Stokes PDEs}\label{secsec:NS}
We consider the 2D in-compressible Navier-Stokes equations as the third benchmark control task, 
\begin{subequations}\label{eq:ns}
    \begin{align}
    &\nabla \cdot \u = 0\,, &\hfill  \label{eq: incompressibility} \\ 
    & \frac{\partial \u}{\partial t} + \u \cdot \nabla \u = -\frac{1}{\rho} \nabla p + \nu \nabla^2 \u & \hfill  \label{eq:momentum} \,.
\end{align}
\end{subequations}
With slight abuse of notation, we denote the spatial variable (in 2D) as $\bm{x} = (x,y) \in \X = [0,1] \times [0,1]$, and $\u = (u, v): \X \times \T \rightarrow \R^2$ represents 2D velocity field,  $\nu$ is the kinematic viscosity of the fluid, $\rho$ is the fluid density, and $p$ is the pressure field. 
Navier-Stokes equation is fundamental in fluid dynamics with extensive applications including aerodynamic design, pollution modeling, and wind turbine flows \cite{Jameson1998, li2021fourier}. For the experiments, we consider boundary control along the top boundary as $\u(x, 0, t) = U(x, t), \forall x \in [0, 1]$.
All other boundary conditions are Dirichlet boundary conditions where the velocity is set to be $0$. The task here is to find boundary control $U(x, t)$ such that the resulting velocity field is close to the reference trajectory given in both desired velocity field $\u_{ref}(\x, t)$ and desired actions $U_{ref}(x, t)$. 
\paragraph{Model-based optimization-based control.} We provide an optimization-based controller based on~\cite{pyta2015optimal} as the model-based control baseline. 
\begin{subequations}\label{eq:optimization_ns}
\begin{eqnarray}
	\min_{U(x, t)}  J(U(\cdot, t), \u) &=& \frac{1}{2} \int_\T \int_\X \|\u(\x,t) - \u_{ref}(\x,t)\|^2 \text{d}\x \text{d} t \nonumber
 \\ && + \frac{\gamma}{2} \int_\T \|U(\cdot, t) -U_{ref}(\cdot, t)\|^2 \text{d} t\\
   \text{s.t.} && \eqref{eq: incompressibility}, \eqref{eq:momentum}, \qquad \u(x, 0, t) = U(x, t), \forall x \in [0, 1].
\end{eqnarray}
\end{subequations}
To ensure computational tractability, the control actions are set to be the tangential, uniform velocity, i.e., $u(x,0,t)=U(t), v(x,0,t)=0$, followed in~\cite{pyta2015optimal}. The optimal control actions are obtained by solving the PDE-constrained optimization problem presented in \eqref{eq:optimization_ns}. This solution employs Lagrange multipliers, utilizing the adjoint method~\cite{mcnamara2004fluid, gunzburger2002perspectives} for gradient computation of the Lagrangian function.

\section{Experiments} 
\label{sec:experiments}
For each of the 3 environments in PDE Control Gym, we implemented baseline model-based control algorithms as well as off-the-shelf RL algorithms including soft actor-critic (SAC)~\citep{haarnoja2018soft} and proximal policy optimization (PPO) \citep{schulman2017proximal} trained using Stable-Baselines3 (Parameters available in Appendix \ref{sec:hyperbolicExample} and \ref{sec:parabolicExample}). We note that all the experiments can be trained in under $1$ hour (Nvidia RTX 3090ti) and entire trajectories can be simulated in seconds.

\begin{table}[ht]
\centering
\begin{tabular}{lc|c|c}
\textbf{}          & \multicolumn{1}{l|}{\textbf{Hyperbolic PDE}}                                                                                 & \multicolumn{1}{l|}{\textbf{Parabolic PDE}}                                                                                  & \multicolumn{1}{l}{\textbf{Navier-Stokes PDE}}                                                                              \\ \cline{2-4} 
\textbf{Algorithm} & \multicolumn{1}{l|}{\textbf{\begin{tabular}[c]{@{}l@{}}Average Episode Reward\\ for Trained Policy $\uparrow$\end{tabular}}} & \multicolumn{1}{l|}{\textbf{\begin{tabular}[c]{@{}l@{}}Average Episode Reward\\ for Trained Policy $\uparrow$\end{tabular}}} & \multicolumn{1}{l}{\textbf{\begin{tabular}[c]{@{}l@{}}Average Episode Reward\\ for Trained Policy $\uparrow$\end{tabular}}} \\ \hline
Model-based        & {\color[HTML]{036400} \textbf{246.3}}                                                                                        & {\color[HTML]{036400} \textbf{299.1}}                                                                                        & -7.931                                                                                                                      \\
PPO                & 172.3                                                                                                                        & 293.3                                                                                                                        & {\color[HTML]{036400} \textbf{-5.370}}                                                                                      \\
SAC                & 184.2                                                                                                                        & 229.1                                                                                                                        & -17.829 \\                                                                        \hline                                           
\end{tabular}
\caption{Resulting control algorithm performance on $50$ test episodes in each gym (larger value indicates better performance). The model-based algorithm for the hyperbolic and parabolic PDEs are the backstepping schemes given in \eqref{eq:bcskhyperbolic}, \eqref{eq:kernel} and \eqref{eq-bkstfbkk}, \eqref{eq-kPDE}, \eqref{eq-kPDEBC0}, \eqref{eq-kPDEBCx} respectively while the method for the Navier-Stokes PDE solves the optimization problem \eqref{eq:optimization_ns}.}
\label{table:mainResults}
\end{table}

\begin{figure}[ht]
    \centering
    \includegraphics[width = 0.95\textwidth, height = 0.3\textwidth]{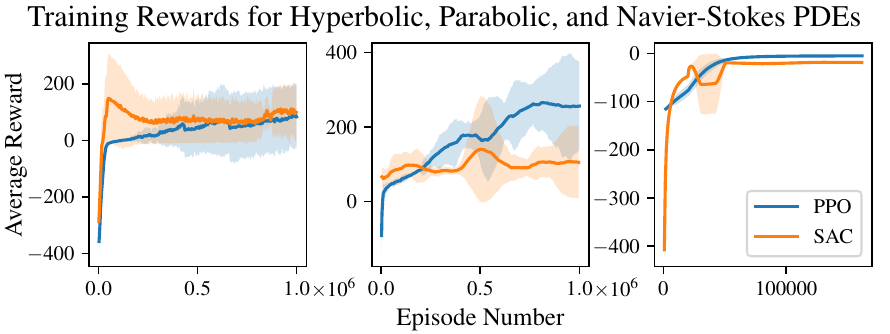}
    \caption{Rewards for training PPO (blue) and SAC (orange) on the 1D transport PDE, 1D reaction-diffusion PDE, and 2D Navier-Stokes PDE from left to right. The solid lines represent the mean and the shaded bounds are $95\%$ confidence intervals across 5 seeds.}
    \label{fig:combinedRewards}
\end{figure}
\subsection{1D Hyperbolic (transport) PDEs} 
\label{sec:hyperbolicExperimental}
\begin{figure}[ht]
    \centering
    \includegraphics{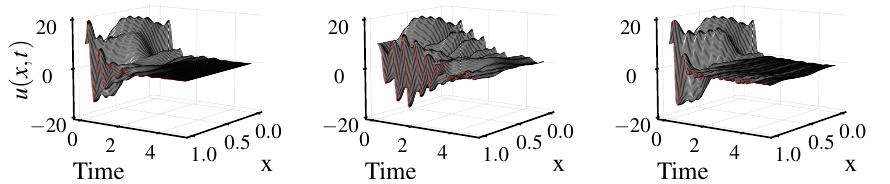}
    \caption{Example of the 1D transport PDE system stabilization using backstepping, PPO, and SAC (left to right) under initial conditions $u(x, 0) = 10$. The recirculation coefficient is defined as $\beta(x)=5\cos(\gamma \cos^{-1}(x))$ with $\gamma=7.35$.}
    \label{fig:hyperbolicTight}
\end{figure}

\paragraph{Experimental design} Our experimental setup for the Hyperbolic 1D problem, detailed in Section \ref{secsec:hyperbolic}, considers full state measurements and boundary control $u(1, t) = U(t)$. We use the Chebyshev polynomial recirculation function $\beta(x)=5\cos(\gamma \cos^{-1}(x))$ from \cite{bhan2023neural}, with $\gamma=7.35$ (future studies may vary $\gamma$). Each episode is initiated from a random initial condition, $u(x, 0) \sim \text{Uniform(1, 10)}$. This setup presents a challenging control scenario, as the open loop system ($U(t)=0$) is unstable (See Figure \ref{fig:hyperbolicOpenloop} in Appendix).

\paragraph{Results}
We now present detailed results on the policies trained and their comparison with model-based backstepping. 
In the left of Figure \ref{fig:combinedRewards}, we present the average reward functions for both RL algorithms over $1$ million training steps. Then, in Table \ref{table:mainResults}, we present the average reward where we run the trained final RL policies, and the model-based backstepping policy for $50$ test episodes with different initial conditions, noting that model-based backstepping performs the best. 
Additionally, in Figures \ref{fig:hyperbolicTight} we provide a comparison across all $3$ control approaches where $u(x, 0)=10$. We can clearly see that although all $3$ policies are stabilizing for the examples, model-based PDE backstepping again performs the best and the RL control signals are high oscillatory leaving room for improvement in applying model-free PDE control algorithms.
\subsection{1D Parabolic (reaction-diffusion) PDEs} 
\begin{figure}[ht]
    \centering
    \includegraphics{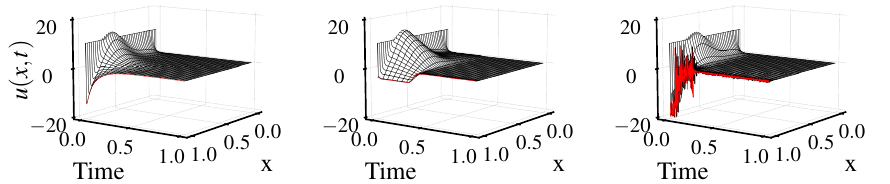}
    \caption{Example of reaction-diffusion PDE system stabilization using backstepping, PPO, and SAC (left to right) under initial conditions $u(x, 0) = 10$. The recirculation coefficient using the Chebyshev polynomial defined as $\lambda(x)=50\cos(\gamma \cos^{-1}(x))$ with $\gamma=8$.}
    \label{fig:parabolicTight}
\end{figure}

\paragraph{Experimental design} 
We adopt the same approach as the 1D Hyperbolic PDE in Section \ref{sec:hyperbolicExperimental} except that the dynamics are now governed by \eqref{eq-PDE}, \eqref{eq-PDEBC0}, with full state measurements and $u(1, t)=U(t)$. The time-horizon is shortened to $1$ second as the algorithms are able to stabilize faster. We choose $\lambda(x) = 50 \cos(\gamma \cos^{-1}(x))$ where $\gamma$ is fixed to be $8$ (future studies may vary $\gamma$). At each episode, the initial conditions are uniformly randomized according to $u(x, 0) \sim \text{Uniform(1, 10)}$, and we note that the system is always open-loop unstable for all possible initial conditions (See Figure \ref{fig:parabolicOpenloop} in Appendix). 
For training, we follow the same procedure as Section \ref{sec:hyperbolicExperimental} except that we require a finer simulation resolution of $\Delta x=0.005$, and the PDE is simulated at $\Delta t=0.00001$ due to the approximation of the second spatial derivative in the reaction-diffusion PDE.

\paragraph{Results}
Figure \ref{fig:combinedRewards} presents the average reward over $1$ million training steps. Unlike the hyperbolic PDE where both RL algorithms performed relatively equal, the PPO algorithm achieved better performance during training which is corroborated by the testing rewards in the middle column of Table \ref{table:mainResults}. 
Figure \ref{fig:parabolicTight} demonstrates a test case with $u(x, 0)=10$. Similar to the transport PDE, we observe oscillations in the RL feedback laws, suggesting potential improvement via enforcing continuity constraints as in \cite{asadi2018lipschitz}. Notably, in Figure \ref{fig:parabolicTight}, perhaps due to reward shaping, PPO differs from the backstepping controller's approach, but maintains excellent performance.  
%
\subsection{2D Navier-Stokes PDEs}
\paragraph{Experimental design} 
For the Navier-Stokes 2D problem (Section~\ref{secsec:NS}), both velocity components are zero initially, i.e., $u(x,y,0)=v(x,y,0)=0$. We apply boundary control on the top boundary with tangential, uniform controlled velocity, setting $u(x,1,t)=U(t)\in \R$ and $v(x,1,t)=0$. For implementation, we discretize the state space with a spatial step of $\Delta x=0.05$ and the PDE is simulated at $\Delta t=0.001$. The reward for training is derived from the negative of the cost in optimization~\eqref{eq:optimization_ns}. The reference velocity vector $\u_{ref}$ is the resulted velocity vector under the boundary control $U(t) = 3 - 5t$, and $U_{ref}=2.0$~.

\begin{figure}[ht]
    \centering
    \includegraphics{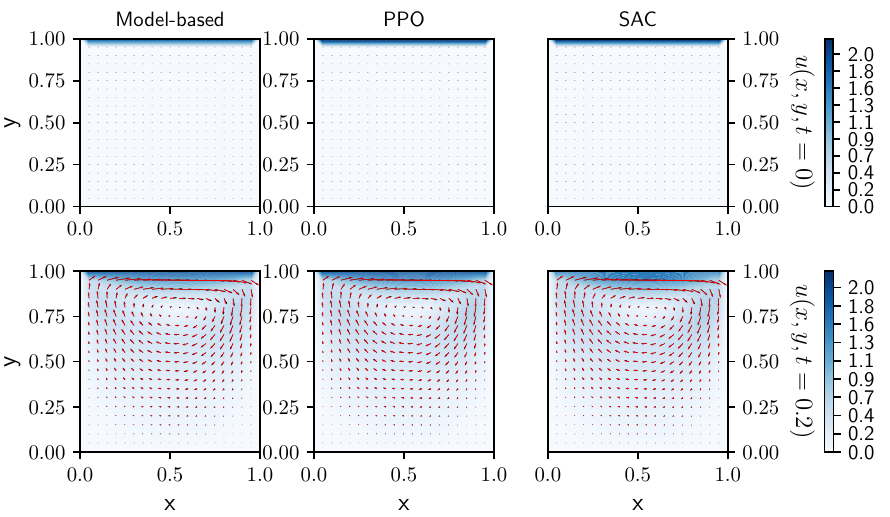}
    \caption{Example of Navier-Stokes PDE tracking using optimization-based control, PPO, and SAC under initial conditions $u(x, y, 0) = 0$ at $t=0$ (top) and $t=0.2$ (bottom). Red and black arrows represent the actual and reference velocity field respectively. The background color represents the magnitude of the velocity vector. }
    \label{fig:nsTight}
\end{figure}

\paragraph{Results}  Figure~\ref{fig:combinedRewards} (right) shows the average reward per episode for PPO and SAC, with PPO outperforming SAC both in terms of higher final rewards and more stable training curves. Table~\ref{table:mainResults} (right) details the average episodic rewards over 50 test episodes, where PPO surpasses both SAC and the model-based optimization algorithm, which often gets stuck in local optima. Despite the reward difference, on a singular example presented in Figure \ref{fig:nsTight}, all methods effectively track the reference velocity vectors. 

\section{Conclusion}
\label{sec:conclusions}
\paragraph{Future work} Throughout this paper, we have mentioned several avenues for future research based on the PDE Control Gym. As such, we conclude this work by briefly summarizing these ideas. 
We employed relatively simple policy network architectures in our RL algorithms, not fully fine-tuning them to the specific problems. The PDE Control Gym presents opportunities to optimize policy network structures, improve reward shaping, and develop better RL algorithms for PDE control tasks. Additionally, our experiments were based on time-invariant linear instability coefficients where $\beta(x)$ and $\lambda(x)$ are unknown but static during RL training. Thus, there is much to be explored for model-free controllers when considering time-varying models, adaptive control, and sensing noise. Furthermore, given the superior performance of backstepping controllers, investigating the potential of pre-training RL methods through imitation learning could be a valuable direction.
\paragraph{Conclusion} In this study, we introduced the first benchmark suite for learning-based boundary control of PDEs. We developed RL gyms for three fundamental PDE control problems: the 1D transport PDE, 1D reaction-diffusion PDE, and 2D Navier Stokes PDE. This gym allows for the separation of algorithm design from the numerical implementation of PDEs. Moreover, we trained a series of \textit{model-free} RL models on the three benchmarks and compared their performance with model-based PDE backstepping and optimization methods. Finally, our work discussed multiple avenues for future research, aiming to inspire new research in the challenging field of PDE control. 

\bibliography{refs.bib}

\begin{thebibliography}{47}
\providecommand{\natexlab}[1]{#1}
\providecommand{\url}[1]{\texttt{#1}}
\expandafter\ifx\csname urlstyle\endcsname\relax
  \providecommand{\doi}[1]{doi: #1}\else
  \providecommand{\doi}{doi: \begingroup \urlstyle{rm}\Url}\fi

\bibitem[Asadi et~al.(2018)Asadi, Misra, and Littman]{asadi2018lipschitz}
Kavosh Asadi, Dipendra Misra, and Michael~L. Littman.
\newblock Lipschitz continuity in model-based reinforcement learning, 2018.

\bibitem[Berberich et~al.(2023)Berberich, Scherer, and Allgöwer]{9903316}
Julian Berberich, Carsten~W. Scherer, and Frank Allgöwer.
\newblock Combining prior knowledge and data for robust controller design.
\newblock \emph{IEEE Transactions on Automatic Control}, 68\penalty0 (8):\penalty0 4618--4633, 2023.
\newblock \doi{10.1109/TAC.2022.3209342}.

\bibitem[Bhan et~al.(2023{\natexlab{a}})Bhan, Shi, and Krstic]{bhan2023neural}
Luke Bhan, Yuanyuan Shi, and Miroslav Krstic.
\newblock Neural operators for bypassing gain and control computations in {PDE} backstepping.
\newblock \emph{IEEE Transactions on Automatic Control}, pages 1--16, 2023{\natexlab{a}}.
\newblock \doi{10.1109/TAC.2023.3347499}.

\bibitem[Bhan et~al.(2023{\natexlab{b}})Bhan, Shi, and Krstic]{pmlr-v211-bhan23a}
Luke Bhan, Yuanyuan Shi, and Miroslav Krstic.
\newblock Operator learning for nonlinear adaptive control.
\newblock In \emph{Proceedings of The 5th Annual Learning for Dynamics and Control Conference (L4DC)}, volume 211 of \emph{Proceedings of Machine Learning Research}, pages 346--357. PMLR, 15--16 Jun 2023{\natexlab{b}}.

\bibitem[Bian et~al.(2024)Bian, Fu, Gupta, and Shi]{bian2024ventilation}
Yuexin Bian, Xiaohan Fu, Rajesh~K Gupta, and Yuanyuan Shi.
\newblock Ventilation and temperature control for energy-efficient and healthy buildings: A differentiable pde approach.
\newblock \emph{arXiv preprint arXiv:2403.08996}, 2024.

\bibitem[Brunke et~al.(2022)Brunke, Greeff, Hall, Yuan, Zhou, Panerati, and Schoellig]{brunke2022safe}
Lukas Brunke, Melissa Greeff, Adam~W Hall, Zhaocong Yuan, Siqi Zhou, Jacopo Panerati, and Angela~P Schoellig.
\newblock Safe learning in robotics: From learning-based control to safe reinforcement learning.
\newblock \emph{Annual Review of Control, Robotics, and Autonomous Systems}, 5:\penalty0 411--444, 2022.

\bibitem[Chen et~al.(2022)Chen, Qu, Tang, Low, and Li]{chen2022reinforcement}
Xin Chen, Guannan Qu, Yujie Tang, Steven Low, and Na~Li.
\newblock Reinforcement learning for selective key applications in power systems: Recent advances and future challenges.
\newblock \emph{IEEE Transactions on Smart Grid}, 13\penalty0 (4):\penalty0 2935--2958, 2022.

\bibitem[Demir et~al.(2024)Demir, Koga, and Krstic]{DEMIR2024111669}
Cenk Demir, Shumon Koga, and Miroslav Krstic.
\newblock Neuron growth control and estimation by pde backstepping.
\newblock \emph{Automatica}, 165:\penalty0 111669, 2024.
\newblock ISSN 0005-1098.
\newblock \doi{https://doi.org/10.1016/j.automatica.2024.111669}.
\newblock URL \url{https://www.sciencedirect.com/science/article/pii/S0005109824001626}.

\bibitem[Di~Meglio et~al.(2012)Di~Meglio, Vazquez, Krstic, and Petit]{6315422}
Florent Di~Meglio, Rafael Vazquez, Miroslav Krstic, and Nicolas Petit.
\newblock Backstepping stabilization of an underactuated 3 × 3 linear hyperbolic system of fluid flow equations.
\newblock In \emph{2012 American Control Conference (ACC)}, pages 3365--3370, 2012.

\bibitem[Duan et~al.(2016)Duan, Chen, Houthooft, Schulman, and Abbeel]{duan2016benchmarking}
Yan Duan, Xi~Chen, Rein Houthooft, John Schulman, and Pieter Abbeel.
\newblock Benchmarking deep reinforcement learning for continuous control.
\newblock In \emph{International conference on machine learning}, pages 1329--1338. PMLR, 2016.

\bibitem[Feng et~al.(2023)Feng, Shi, Qu, Low, Anandkumar, and Wierman]{10336939}
Jie Feng, Yuanyuan Shi, Guannan Qu, Steven~H. Low, Anima Anandkumar, and Adam Wierman.
\newblock Stability constrained reinforcement learning for decentralized real-time voltage control.
\newblock \emph{IEEE Transactions on Control of Network Systems}, pages 1--12, 2023.
\newblock \doi{10.1109/TCNS.2023.3338240}.

\bibitem[Gunzburger(2002)]{gunzburger2002perspectives}
Max~D Gunzburger.
\newblock \emph{Perspectives in flow control and optimization}.
\newblock SIAM, 2002.

\bibitem[Gupta and Brandstetter(2022)]{gupta2022towards}
Jayesh~K Gupta and Johannes Brandstetter.
\newblock Towards multi-spatiotemporal-scale generalized {PDE} modeling.
\newblock \emph{{\em arXiv:2209.15616}}, 2022.

\bibitem[Haarnoja et~al.(2018{\natexlab{a}})Haarnoja, Zhou, Abbeel, and Levine]{haarnoja2018soft}
Tuomas Haarnoja, Aurick Zhou, Pieter Abbeel, and Sergey Levine.
\newblock Soft actor-critic: Off-policy maximum entropy deep reinforcement learning with a stochastic actor.
\newblock In \emph{International conference on machine learning}, pages 1861--1870. PMLR, 2018{\natexlab{a}}.

\bibitem[Haarnoja et~al.(2018{\natexlab{b}})Haarnoja, Zhou, Abbeel, and Levine]{pmlr-v80-haarnoja18b}
Tuomas Haarnoja, Aurick Zhou, Pieter Abbeel, and Sergey Levine.
\newblock Soft actor-critic: Off-policy maximum entropy deep reinforcement learning with a stochastic actor.
\newblock In Jennifer Dy and Andreas Krause, editors, \emph{Proceedings of the 35th International Conference on Machine Learning}, volume~80 of \emph{Proceedings of Machine Learning Research}, pages 1861--1870. PMLR, 10--15 Jul 2018{\natexlab{b}}.
\newblock URL \url{https://proceedings.mlr.press/v80/haarnoja18b.html}.

\bibitem[Holl et~al.(2020)Holl, Thuerey, and Koltun]{Holl2020Learning}
Philipp Holl, Nils Thuerey, and Vladlen Koltun.
\newblock Learning to control {PDE}s with differentiable physics.
\newblock In \emph{International Conference on Learning Representations (ICLR)}, 2020.

\bibitem[Huan~Yu(2023)]{trafficHuanYu}
Miroslav~Krstic Huan~Yu.
\newblock \emph{Traffic Congestion Control by PDE Backstepping}.
\newblock Birkhäuser Cham, 2023.

\bibitem[Izadi et~al.(2015)Izadi, Abdollahi, and Dubljevic]{IZADI201541}
Mojtaba Izadi, Javad Abdollahi, and Stevan~S. Dubljevic.
\newblock {PDE} backstepping control of one-dimensional heat equation with time-varying domain.
\newblock \emph{Automatica}, 54:\penalty0 41--48, 2015.

\bibitem[Jameson et~al.(1998)Jameson, Martinelli, and Pierce]{Jameson1998}
A.~Jameson, L.~Martinelli, and N.~A. Pierce.
\newblock Optimum aerodynamic design using the {Navier-Stokes} equations.
\newblock \emph{Theoretical and Computational Fluid Dynamics}, 10\penalty0 (1):\penalty0 213--237, Jan 1998.

\bibitem[Kiran et~al.(2021)Kiran, Sobh, Talpaert, Mannion, Al~Sallab, Yogamani, and P{\'e}rez]{kiran2021deep}
B~Ravi Kiran, Ibrahim Sobh, Victor Talpaert, Patrick Mannion, Ahmad~A Al~Sallab, Senthil Yogamani, and Patrick P{\'e}rez.
\newblock Deep reinforcement learning for autonomous driving: A survey.
\newblock \emph{IEEE Transactions on Intelligent Transportation Systems}, 23\penalty0 (6):\penalty0 4909--4926, 2021.

\bibitem[Koga and Krstic(2020)]{koga}
Shumon Koga and Miroslav Krstic.
\newblock \emph{Materials Phase Change {PDE} Control \& Estimation}.
\newblock Birkh\"{a}user, 2020.

\bibitem[Krizhevsky et~al.(2012)Krizhevsky, Sutskever, and Hinton]{NIPS2012_c399862d}
Alex Krizhevsky, Ilya Sutskever, and Geoffrey~E Hinton.
\newblock Imagenet classification with deep convolutional neural networks.
\newblock In F.~Pereira, C.J. Burges, L.~Bottou, and K.Q. Weinberger, editors, \emph{Advances in Neural Information Processing Systems}, volume~25, 2012.

\bibitem[Krstic and Smyshlyaev(2008{\natexlab{a}})]{KRSTIC2008750}
Miroslav Krstic and Andrey Smyshlyaev.
\newblock Backstepping boundary control for first-order hyperbolic {PDE}s and application to systems with actuator and sensor delays.
\newblock \emph{Systems \& Control Letters}, 57\penalty0 (9):\penalty0 750--758, 2008{\natexlab{a}}.

\bibitem[Krstic and Smyshlyaev(2008{\natexlab{b}})]{doi:10.1137/1.9780898718607}
Miroslav Krstic and Andrey Smyshlyaev.
\newblock \emph{Boundary Control of PDEs}.
\newblock Society for Industrial and Applied Mathematics, Philadelphia, PA, 2008{\natexlab{b}}.

\bibitem[Krstic et~al.(2024)Krstic, Bhan, and Shi]{krstic2023neural}
Miroslav Krstic, Luke Bhan, and Yuanyuan Shi.
\newblock Neural operators of backstepping controller and observer gain functions for reaction–diffusion {PDE}s.
\newblock \emph{Automatica}, 164:\penalty0 111649, 2024.
\newblock ISSN 0005-1098.
\newblock \doi{https://doi.org/10.1016/j.automatica.2024.111649}.
\newblock URL \url{https://www.sciencedirect.com/science/article/pii/S0005109824001420}.

\bibitem[Le and Moin(1991)]{le1991improvement}
Hung Le and Parviz Moin.
\newblock An improvement of fractional step methods for the incompressible {Navier-Stokes} equations.
\newblock \emph{Journal of computational physics}, 92\penalty0 (2):\penalty0 369--379, 1991.

\bibitem[LeVeque(1992)]{books/daglib/0078096}
Randall~J. LeVeque.
\newblock \emph{Numerical methods for conservation laws (2. ed.).}
\newblock Lectures in mathematics. Birkhäuser, 1992.
\newblock ISBN 978-3-7643-2723-1.

\bibitem[Li et~al.(2021)Li, Kovachki, Azizzadenesheli, liu, Bhattacharya, Stuart, and Anandkumar]{li2021fourier}
Zongyi Li, Nikola~Borislavov Kovachki, Kamyar Azizzadenesheli, Burigede liu, Kaushik Bhattacharya, Andrew Stuart, and Anima Anandkumar.
\newblock Fourier neural operator for parametric partial differential equations.
\newblock In \emph{International Conference on Learning Representations (ICLR)}, 2021.

\bibitem[Lu et~al.(2021)Lu, Jin, Pang, Zhang, and Karniadakis]{lu2021}
Lu~Lu, Pengzhan Jin, Guofei Pang, Zhongqiang Zhang, and George~Em Karniadakis.
\newblock Learning nonlinear operators via {DeepONet} based on the universal approximation theorem of operators.
\newblock \emph{Nature Machine Intelligence}, 3\penalty0 (3):\penalty0 218--229, 2021.

\bibitem[McNamara et~al.(2004)McNamara, Treuille, Popovi{\'c}, and Stam]{mcnamara2004fluid}
Antoine McNamara, Adrien Treuille, Zoran Popovi{\'c}, and Jos Stam.
\newblock Fluid control using the adjoint method.
\newblock \emph{ACM Transactions On Graphics (TOG)}, 23\penalty0 (3):\penalty0 449--456, 2004.

\bibitem[Mildenhall et~al.(2020)Mildenhall, Srinivasan, Tancik, Barron, Ramamoorthi, and Ng]{mildenhall2020nerf}
Ben Mildenhall, Pratul~P. Srinivasan, Matthew Tancik, Jonathan~T. Barron, Ravi Ramamoorthi, and Ren Ng.
\newblock Nerf: Representing scenes as neural radiance fields for view synthesis.
\newblock In \emph{ECCV}, 2020.

\bibitem[Moura et~al.(2014)Moura, Chaturvedi, and Krsti{\'c}]{moura2014adaptive}
Scott~J Moura, Nalin~A Chaturvedi, and Miroslav Krsti{\'c}.
\newblock Adaptive partial differential equation observer for battery state-of-charge/state-of-health estimation via an electrochemical model.
\newblock \emph{Journal of Dynamic Systems, Measurement, and Control}, 136\penalty0 (1):\penalty0 011015, 2014.

\bibitem[Mowlavi and Nabi(2023)]{MOWLAVI2023111731}
Saviz Mowlavi and Saleh Nabi.
\newblock Optimal control of {PDE}s using physics-informed neural networks.
\newblock \emph{Journal of Computational Physics}, 473:\penalty0 111731, 2023.

\bibitem[Pyta et~al.(2015)Pyta, Herty, and Abel]{pyta2015optimal}
Lorenz Pyta, Michael Herty, and Dirk Abel.
\newblock Optimal feedback control of the incompressible {Navier-Stokes}-equations using reduced order models.
\newblock In \emph{2015 54th IEEE Conference on Decision and Control (CDC)}, pages 2519--2524. IEEE, 2015.

\bibitem[Qi et~al.(2023)Qi, Zhang, and Krstic]{qi2023neural}
Jie Qi, Jing Zhang, and Miroslav Krstic.
\newblock Neural operators for delay-compensating control of hyperbolic {PIDE}s.
\newblock \emph{{\em arXiv:2307.11436}}, 2023.

\bibitem[Raffin et~al.(2021)Raffin, Hill, Gleave, Kanervisto, Ernestus, and Dormann]{raffin2021stablebaselines3}
Antonin Raffin, Ashley Hill, Adam Gleave, Anssi Kanervisto, Maximilian Ernestus, and Noah Dormann.
\newblock Stable-{Baselines3}: {Reliable} {Reinforcement} {Learning} {Implementations}.
\newblock \emph{Journal of Machine Learning Research}, 22\penalty0 (268):\penalty0 1--8, 2021.

\bibitem[Raissi et~al.(2019)Raissi, Perdikaris, and Karniadakis]{RAISSI2019686}
M.~Raissi, P.~Perdikaris, and G.E. Karniadakis.
\newblock Physics-informed neural networks: A deep learning framework for solving forward and inverse problems involving nonlinear partial differential equations.
\newblock \emph{Journal of Computational Physics}, 378:\penalty0 686--707, 2019.

\bibitem[Schulman et~al.(2017)Schulman, Wolski, Dhariwal, Radford, and Klimov]{schulman2017proximal}
John Schulman, Filip Wolski, Prafulla Dhariwal, Alec Radford, and Oleg Klimov.
\newblock Proximal policy optimization algorithms.
\newblock \emph{{\em arXiv:1707.06347}}, 2017.

\bibitem[Shi et~al.(2022)Shi, Li, Yu, Steeves, Anandkumar, and Krstic]{9992759}
Yuanyuan Shi, Zongyi Li, Huan Yu, Drew Steeves, Anima Anandkumar, and Miroslav Krstic.
\newblock Machine learning accelerated {PDE} backstepping observers.
\newblock In \emph{2022 IEEE 61st Conference on Decision and Control (CDC)}, pages 5423--5428, 2022.

\bibitem[Smyshlyaev and Krstic(2004)]{1369395}
A.~Smyshlyaev and M.~Krstic.
\newblock Closed-form boundary state feedbacks for a class of 1-{D} partial integro-differential equations.
\newblock \emph{IEEE Transactions on Automatic Control}, 49\penalty0 (12):\penalty0 2185--2202, 2004.

\bibitem[Smyshlyaev and Krstic(2010)]{Smyshlyaev2010}
A.~Smyshlyaev and M.~Krstic.
\newblock \emph{Adaptive Control of Parabolic {PDE}s}.
\newblock Princeton University Press, 2010.

\bibitem[Takamoto et~al.(2022)Takamoto, Praditia, Leiteritz, MacKinlay, Alesiani, Pfl\"{u}ger, and Niepert]{NEURIPS2022_0a974713}
Makoto Takamoto, Timothy Praditia, Raphael Leiteritz, Daniel MacKinlay, Francesco Alesiani, Dirk Pfl\"{u}ger, and Mathias Niepert.
\newblock {PDEBench:} an extensive benchmark for scientific machine learning.
\newblock In S.~Koyejo, S.~Mohamed, A.~Agarwal, D.~Belgrave, K.~Cho, and A.~Oh, editors, \emph{Advances in Neural Information Processing Systems}, volume~35, pages 1596--1611, 2022.

\bibitem[Tassa et~al.(2018)Tassa, Doron, Muldal, Erez, Li, de~Las~Casas, Budden, Abdolmaleki, Merel, Lefrancq, Lillicrap, and Riedmiller]{DBLP:journals/corr/abs-1801-00690}
Yuval Tassa, Yotam Doron, Alistair Muldal, Tom Erez, Yazhe Li, Diego de~Las~Casas, David Budden, Abbas Abdolmaleki, Josh Merel, Andrew Lefrancq, Timothy~P. Lillicrap, and Martin~A. Riedmiller.
\newblock Deepmind control suite.
\newblock \emph{CoRR}, abs/1801.00690, 2018.

\bibitem[Vazquez et~al.(2023)Vazquez, Chen, Qiao, and Krstic]{10384080}
Rafael Vazquez, Guangwei Chen, Junfei Qiao, and Miroslav Krstic.
\newblock The power series method to compute backstepping kernel gains: Theory and practice.
\newblock In \emph{2023 62nd IEEE Conference on Decision and Control (CDC)}, pages 8162--8169, 2023.
\newblock \doi{10.1109/CDC49753.2023.10384080}.

\bibitem[Wang et~al.(2020)Wang, Wang, and Xu]{wang2020modeling}
Haiyan Wang, Feng Wang, and Kuai Xu.
\newblock \emph{Modeling information diffusion in online social networks with partial differential equations}, volume~7.
\newblock Springer Nature, 2020.

\bibitem[Yu et~al.(2022)Yu, Park, Bayen, Moura, and Krstic]{9568241}
Huan Yu, Saehong Park, Alexandre Bayen, Scott Moura, and Miroslav Krstic.
\newblock Reinforcement learning versus {PDE} backstepping and {PI} control for congested freeway traffic.
\newblock \emph{IEEE Transactions on Control Systems Technology}, 30\penalty0 (4):\penalty0 1595--1611, 2022.

\bibitem[Zhang et~al.(2024)Zhang, Mao, Mowlavi, Benosman, and Başar]{zhang2024controlgym}
Xiangyuan Zhang, Weichao Mao, Saviz Mowlavi, Mouhacine Benosman, and Tamer Başar.
\newblock Controlgym: Large-scale control environments for benchmarking reinforcement learning algorithms, 2024.

\end{thebibliography}







\newpage 
\appendix
\section{Hyperbolic Partial Differential Equation}
\subsection{First-Order Finite Difference Scheme}\label{sec:hyperbolicFiniteDiff}
We consider the benchmark transport PDE in the form 
\begin{eqnarray} \label{eq:appendixHyperbolic}
    u_t &=& u_x + \beta(x)u(0, t), \quad (x, t) \in [0, 1) \times \mathbb{R}_+ \,, 
\end{eqnarray}
with a single boundary condition representing the control input in either Dirichlet Neumann form at $x=1$ ($u(1, t) = U(t)$). Notice that instability is caused by the recirculation term $\beta(x)u(0,t)$ - otherwise the resulting PDE with be considered an instantiation of the inviscid Burger's equation. For the numerical scheme, consider the first-order Taylor approximation for $u$ and the resulting substitution of derivatives yields
\begin{eqnarray}
    u_j^{n+1} = u_j^n + \Delta t \left(\frac{u_{j+1}^n - u_{j}^n}{\Delta x} + \beta_j u_0^{n}\right)\,, 
\end{eqnarray}
where $\Delta  t$ denotes the temporal timestep, $\Delta x$ denotes the spatial timestep, $n=0,...,Nt$, $j=0,...,Nx$ where $Nt$ and $Nx$ are the total number of temporal and spatial steps respectively. 
To enforce Neumann boundary conditions, let $u_\zeta^n|_{\zeta=Nx}$ represent the spatial derivative at time $t$ of spatial  point $Nx$ which is given by the user as control input. Then, we have 
\begin{eqnarray} \label{eq:hyperbolicNeumann}
    u_\zeta^n|_{\zeta=Nx} = \frac{u_{Nx}^{n} - u_{Nx-1}^n}{\Delta x}\,,
\end{eqnarray}
which is rearranged for the final boundary point
\begin{eqnarray}\label{eq:hyperNeumann}
    u_{Nx}^n = u_{Nx-1}^n + (\Delta x) u_\zeta^n|_{\zeta=Nx}\,.
\end{eqnarray}
In the case of Dirichlet boundary conditions, the computation is straightforward as $u_{Nx}^n$ is directly set as the given control input. 
For stabilization of the finite-difference scheme, it is recommended that timestep be much much smaller than the spatial step \cite{books/daglib/0078096}.
\newpage

\newpage
\subsection{Details on Numerical Implementations for Benchmark 1D Hyperoblic PDE}\label{sec:hyperbolicExample}

\subsubsection{Reinforcement learning baselines: Hyperparameters for proximal policy optimization (PPO) and soft-actor critic (SAC)} \label{sec:hyperbolicRLParams}
We provide the entire set of hyperparameters for training the PPO and SAC algorithms in Table \ref{table:ppoParameters} and \ref{table:sacParameters} respectively. For a details on the RL learning algorithms, see \cite{schulman2017proximal} (PPO) and \cite{pmlr-v80-haarnoja18b} (SAC). 
\begin{multicols}{2}

\begin{table}[H]
\begin{tabular}{l|l}
PPO Parameter                                                                                    & Value  \\ \hline
Learning Rate                                                                                    & 0.0003 \\ \hline
\begin{tabular}[c]{@{}l@{}}Num Steps per\\ Update\end{tabular}                                   & 2048   \\ \hline
Batch\_size                                                                                      & 64     \\ \hline
\begin{tabular}[c]{@{}l@{}}Num Epcohs\\ per Surrogate Loss \\ Update\end{tabular}                & 10     \\ \hline
Discount Factor $\gamma$                                                                         & 0.99   \\ \hline
\begin{tabular}[c]{@{}l@{}}Bias vs Variance \\ Trade-Off for \\ Advantage Estimator\end{tabular} & 0.95   \\ \hline
Clipping Parameter $\epsilon$                                                                    & 0.2    \\ \hline
Entropy Coefficient for Loss                                                                     & 0.0    \\ \hline
\begin{tabular}[c]{@{}l@{}}Value Function Coefficient\\ for Loss\end{tabular}                    & 0.5    \\ \hline
\begin{tabular}[c]{@{}l@{}}Max Value for \\ Gradient Clipping\end{tabular}                       & 0.5   
\end{tabular}
\caption{Parameters for PPO Model Trained}
\label{table:ppoParameters}
\end{table}

\begin{table}[H]
\centering
\begin{tabular}{l|l}
SAC Parameter                  & Value   \\ \hline
Learning Rate                  & 0.0003  \\ \hline
Buffer Size                    & 1000000 \\ \hline
Batch Size                     & 256     \\ \hline
Soft Update Coefficient $\tau$ & 0.005   \\ \hline
Discount factor $\gamma$       & 0.99    \\ \hline
Action Noise                   & None   
\end{tabular}
\caption{Parameters for SAC Model Trained}
\label{table:sacParameters}
\end{table}

\end{multicols}

\begin{table}[ht]
\centering
\resizebox{\textwidth}{!}{
\begin{tabular}{l|lll}
\begin{tabular}[c]{@{}l@{}}Markov Decision \\ Process Tuple\end{tabular}                                  & \begin{tabular}[c]{@{}l@{}}Parameter for Hyperbolic\\ PDE Benchmark Example\end{tabular}                                                                                                                                                                                                  &  &  \\ \cline{1-2}
$\mathcal{S}$ - state space                                                                               & $\mathcal{S} \subseteq C[0, 1]$                                                                                                                                                                                                                                                           &  &  \\
$\mathcal{A}$ - action space                                                                              & $\mathcal{A} \subseteq \{x \in \mathbb{R}: -20 \leq x \leq 20 \} $                                                                                                                        &  &  \\
$\mathcal{O}$ - observation space                                                                         & \begin{tabular}[c]{@{}l@{}}$\mathcal{O} = \mathcal{S}$\\  (can be varied to user choice)\end{tabular}                                                                                                                                                                          &  &  \\
\begin{tabular}[c]{@{}l@{}}$p_0$ - initial sample \\ pdf for $u(x, 0) \in \mathcal{S} $\end{tabular}                       & \begin{tabular}[c]{@{}l@{}}$p_0 \subseteq \{f \in C[0, 1] | f(x) = c, \quad \forall x \in [0, 1] \}$\\ where $c \sim \text{Uniform}(1, 10)$\end{tabular}                                                                                                                                               &  &  \\
$p_f(\cdot | x_t, a_t)$ - state transition                                                                & \begin{tabular}[c]{@{}l@{}}Dynamics described by PDE in \eqref{eq:hyperbolic}\\  with $u(1, t)=a_t$\end{tabular}                                                                                                                                                         &  &  \\
$T$ - time horizon                                                                                        & $5$ seconds                                                                                                                                                                                                                                                                               &  &  \\
$r_a(s, s')$ - reward ($s \in \mathcal{S}$)                                                               & $r_a(s, s') = -1*\|s'-s\|_{L_2}$                                                                                                                                                                                                                                                          &  &  \\
\begin{tabular}[c]{@{}l@{}}$q(s_T, a_0, ..., a_T) $ \\ terminal cost ($s_T \in \mathcal{S}$)\end{tabular} & \begin{tabular}[c]{@{}l@{}}$q(s_T, a_0, ..., a_T) = \begin{cases}\\ 0 & \|s_T\|_{L_2} > \zeta \\\\ \sigma - 1/\eta * \sum_{\tau=0}^T|a_\tau|_{L_1} - \|s_t\|_{L_2}& \|s_T\|_{L_2} \leq \zeta \end{cases}$\\ where $\sigma, \eta, \zeta$ are hyperparameters\end{tabular} &  & 
\end{tabular}
}
\caption{Markov Decision Process describing the PDE example in Section \ref{sec:hyperbolicExample}}
\label{table:hyperbolicMDP}
\end{table}
\subsubsection{Experimental Design for 1D Hyperbolic PDE}
\begin{figure}[ht]
    \centering
    \includegraphics{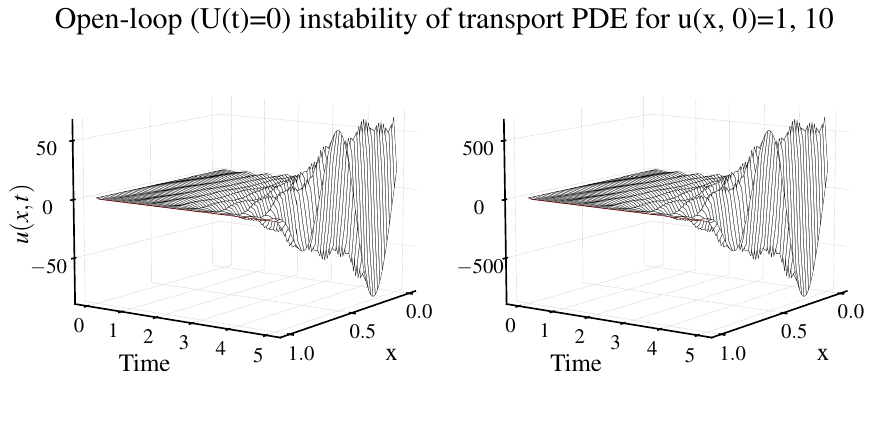}
    \caption{Instability of the 1D transport PDE with $\beta(x) = 5\cos(7.35\cos^{-1}(x))$ under a openloop control signal ($U(t)=0$).}
    \label{fig:hyperbolicOpenloop}
\end{figure}

\begin{figure}[H]
    \centering
    \includegraphics{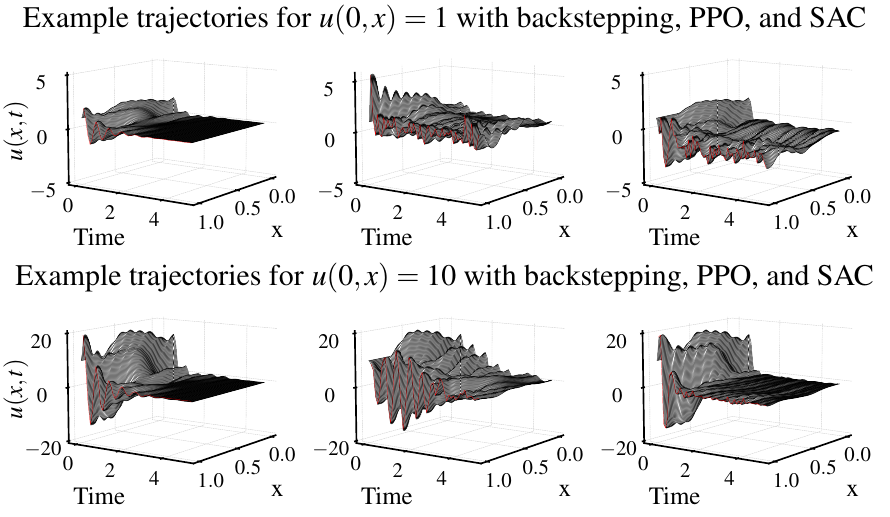}
    \caption{Examples of PDE system stabilization using backstepping, PPO, and SAC (left to right) under two different initial conditions $u(x, 0) = 1, 10$. The PDE has functional recirculation coefficient using the Chebyhsev polynomial defined as $\beta(x)=5\cos(\gamma \cos^{-1}(x))$ with $\gamma=7.35$. The control signal for each approach is marked in red and in Figure \ref{fig:hyperbolicControlSignals}.}
    \label{fig:hyperbolicExamples}
\end{figure}

We now discuss the construction of each episode for reinforcement learning which is concurrently summarized as a Markov Decision Process (MDP) in Table \ref{table:hyperbolicMDP}. For this work, we consider a simplified version of the PDE where $\gamma$ is fixed for the entire training process to $\gamma=7.35$. Then, each episode randomly begins with an initial condition sampled from $u(x, 0) \sim \text{Uniform}(1, 10)$. This creates a sufficiently challenging PDE to control as for any initial condition, the PDE is open loop unstable. As an example, we present the PDEs for $u(x, 0)=1, 10$ in the left and right of Figure \ref{fig:hyperbolicOpenloop}. 
\newpage 
For the RL policies, we utilized a novel reward function designed to accurately handle the challenging oscillations of hyperbolic PDEs (See Figure \ref{fig:hyperbolicExamples} with the backstepping controller for an example). With this, we developed the following reward function for 1D PDEs: 
\begin{eqnarray}
    r_a(s, s') &=& -1*\|s'-s\|_{L_2} \,,\\ 
    q(s_T, a_0, ..., a_T) &=& \begin{cases}\\ 0 & \|s_T\|_{L_2} > 20 \\\\ \sigma - 1/\eta * \sum_{\tau=0}^T|a_\tau|_{L_1} - \|s_t\|_{L_2}& \|s_T\|_{L_2} \leq 20 \end{cases}\,,
\end{eqnarray}
where $\sigma, \eta, \zeta$ were hyperparameters set to $300, 1000, 20$. The intuition of this reward function is in two components. First, $r_a(s, s')$ rewards control inputs that force the state to become smaller compared to its previous state leading to an encouragement towards stabilization. Then, at the termination of the episode, if the final state $L_2$ norm is small enough ($\leq 20$), we give the policy a large reward which is now penalized by the control costs - ie, we only value control costs for the policy after we know that the policy for the episode is stabilizing. This trade-off is commonly found in a series of optimal control problems - but due to the challenging nature of PDEs, one is usually satisfied with just stabilization guarantees without limits on the control inputs. This highlights one of the advantages of using model-free approaches where one can begin to enforce a stabilization first, minimization of control second approach. Lastly, we note that many rewards will encourage successful policies and we leave it to the community to develop better tuned rewards for these benchmark PDE problems.

For training, we run each model with parameters listed in Tables \ref{table:ppoParameters} and \ref{table:sacParameters} (standard defaults for stable-baselines3, non-optimized) in the environment governed by the MDP in Table \ref{table:hyperbolicMDP}. From an implementation perspective, we discretize both the state space using a spatial step of $dx=0.01$ and further discretize each timestep of the MDP by $dt=0.01$ as well. However, we note that the first-order PDE scheme in Section \ref{sec:hyperbolicFiniteDiff} requires a much smaller timestep then spatial step and thus, the PDE is internally simulated at a timestep of $dt_{PDE}=0.0001$ where it receives a new control input at every $0.01$ second and maintains the same input for the higher resolution PDE simulation until the next $0.01$ second is reached. Naturally, it is worth considering sampling of the controller at the same rate of the PDE - but this results in poor performance for the reinforcement learning models and is less indicative of a real-world application given it is challenging to provide control inputs at frequencies larger than $100$Hz. 
\newpage 
\subsection{Case Study: Hyperbolic PDE}
In this section, we show the results for two individual initial conditions for the Hyperbolic PDE problem, namely $u(x, 0) = 0$ and $u(x, 0) = 10$.

\begin{table}[H]
\centering
\begin{tabular}{c|c|c|c}
\multicolumn{1}{l|}{\textbf{Initial Condition}} & \multicolumn{1}{l|}{\textbf{Control Algorithm}} & \multicolumn{1}{l|}{\textbf{\begin{tabular}[c]{@{}l@{}}Total Episode \\ Reward $\uparrow$\end{tabular}}} & \multicolumn{1}{l}{\textbf{\begin{tabular}[c]{@{}l@{}}Total Episode State \\ Summed $L_2$ Norm\\ ($\sum_{t=0}^T \|u(\cdot, t) \|_{L_2}$) $\downarrow$\end{tabular}}} \\ \hline
$u(x, 0)=1$                                     & Backstepping                                    & {\color[HTML]{036400} \textbf{289.8}}                                                                    & {\color[HTML]{036400} \textbf{106.1}}                                                                                                                          \\
$u(x, 0)=1$                                     & PPO                                             & 246.0                                                                                                    & 448.1                                                                                                                                                          \\
$u(x, 0)=1$                                     & SAC                                             & 212.9                                                                                                    & 720.4                                                                                                                                                          \\
$u(x, 0)=10$                                    & Backstepping                                    & {\color[HTML]{036400} \textbf{198.4}}                                                                    & {\color[HTML]{036400} \textbf{1060.9}}                                                                                                                         \\
$u(x, 0)=10$                                    & PPO                                             & 32.7                                                                                                     & 2026.4                                                                                                                                                         \\
$u(x, 0)=10$                                    & SAC                                             & 133.7                                                                                                    & 1402.8                                                                                                                                                        
\end{tabular}%
\caption{Resulting control performance for backstepping, PPO, and SAC on the two PDE examples corresponding to Figure \ref{fig:hyperbolicExamples}.}
\label{table:hyperbolicPerformance}
\end{table}

\vspace{-10pt}
\begin{figure}[H]
    \centering
    \includegraphics{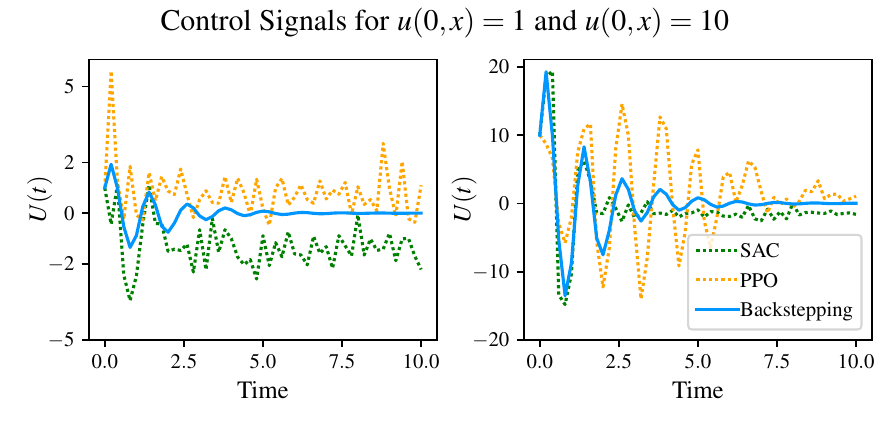}
    \caption{The corresponding control signal given to the PDEs in Figure \ref{fig:hyperbolicExamples}. The control signals on the left match the PDE with $u(x, 0)=1$ (top-row of Fig. \ref{fig:hyperbolicExamples}) and the signals on the right correspond to the PDE with $u(x, 0)=10$ (bottom-row of Fig. \ref{fig:hyperbolicExamples}) respectively.}
    \label{fig:hyperbolicControlSignals}
\end{figure}

\section{Reaction-Diffusion Partial Differential Equation}\label{sec:parabolic}
\subsection{First-Order Finite Difference Scheme}\label{sec:parabolicFiniteDiff}
We consider the benchmark reaction-diffusion PDE in the form 
\begin{eqnarray}\label{eq:appendixparabolic}
    u_t &=& u_{xx} + \lambda(x)u(x, t), \quad (x, t) \in (0, 1) \times \mathbb{R}_+\,,
    \\
    u(0, t) &=& 0 \label{eq:appendixparabolicBc}\,,
\end{eqnarray}
with the control input again in either Dirichlet or Neumann form at $x=1$ ($u(1, t) = U(t)$). As with the transport PDE, the recirculation term $\lambda(x)u(x, t)$ causes instability as otherwise the resulting PDE would reduce to the stable heat equation. For the numerical scheme, consider the first-order Taylor approximation for $u$ as
\begin{eqnarray}
    u_j^{n+1} = u_j^n +  \Delta t \left(\frac{u_{j-1}^n - 2 u_j^n + u_{j+1}^n}{(\Delta x)^2} + \lambda_j u_j^n \right) \,,
\end{eqnarray}
where $\Delta t$ denotes the temporal timestep, $\Delta x$ denotes the spatial timestep, $n=0,..., Nt$, $j=0,...,Nx$ where $Nt$ and $Nx$ are the total number of time steps respectively. Explicitly we set $u_0^k=0 \forall k\in[0, Nt]$ for the first boundary condition and the second boundary condition follows as in the Hyperbolic PDE case with \eqref{eq:hyperNeumann} for Neumann boundary conditions and $u_{Nx}^n$ for Dirichlet boundary conditions. Again, we require extremely small spatial and temporal time-steps for stabilization of the scheme.

\subsection{Details on numerical implementation for benchmark 1D reaction-diffusion PDE}\label{sec:parabolicExample}
\begin{figure}[ht]
    \centering
    \includegraphics{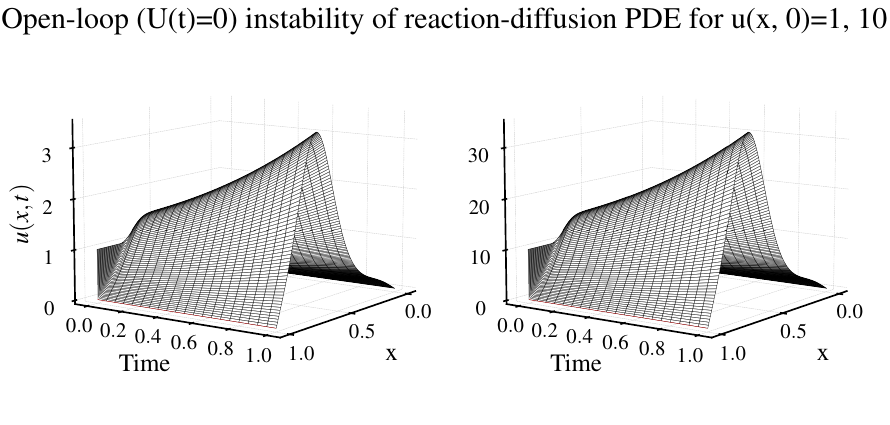}
    \caption{Instability of the reaction-diffusion with $\beta(x) = 50\cos(8\cos^{-1}(x))$ under a openloop control signal ($U(t)=0$).}
    \label{fig:parabolicOpenloop}
\end{figure}

We adopt the exact same, \textbf{untuned}, hyperparameters for PPO and SAC as in Section \ref{sec:hyperbolicRLParams} and adopt the same MDP as presented in Table \ref{table:hyperbolicMDP} except that the dynamics are now governed by \eqref{eq:appendixparabolic}, \eqref{eq:appendixparabolicBc}, with $u(1, t)=a_t$ and the time-horizon is shortened to $1$ second as the algorithms are able to stabilize faster. We choose $\lambda(x) = 50 \cos(\gamma \cos^{-1}(x))$ where $\gamma$ is fixed to be $8$. With the following configuration and initial conditions again sampled between $1$ and $10$, we see that the PDE is openloop unstable in Figure \ref{fig:parabolicOpenloop}. 

For training, we follow the same procedure as Section \ref{sec:hyperbolicExample} except that we require a finer simulation resolution of $dt=0.001$, $dx=0.005$, and $dt_{PDE}=0.00001$ due to the approximation of the second spatial derivative in the reaction diffusion PDE. 
\newpage 
\subsubsection{Case Study: Parabolic PDE}
In this section, we show the results for two individual initial conditions for the parabolic PDE problem, namely $u(x, 0) = 0$ and $u(x, 0) = 10$.
\begin{table}[H]
\centering
\resizebox{\textwidth}{!}{%
\begin{tabular}{c|c|c|c}
\multicolumn{1}{l|}{\textbf{Initial Condition}} & \multicolumn{1}{l|}{\textbf{Control Algorithm}} & \multicolumn{1}{l|}{\textbf{\begin{tabular}[c]{@{}l@{}}Total Episode \\ Reward $\uparrow$\end{tabular}}} & \multicolumn{1}{l}{\textbf{\begin{tabular}[c]{@{}l@{}}Total Episode \\ Summed $L_2$ Norm\\ ($\sum_{t=0}^T \|u(\cdot, t) \|_{L_2}$) $\downarrow$\end{tabular}}} \\ \hline
$u(x, 0)=1$                                     & Backstepping                                    & {\color[HTML]{036400} \textbf{299.8}}                                                                    & {\color[HTML]{000000} 1275.4}                                                                                                                                  \\
$u(x, 0)=1$                                     & PPO                                             & 295.1                                                                                                    & {\color[HTML]{036400} \textbf{1103.5}}                                                                                                                         \\
$u(x, 0)=1$                                     & SAC                                             & 239.8                                                                                                    & 1968.7                                                                                                                                                         \\
$u(x, 0)=10$                                    & Backstepping                                    & {\color[HTML]{036400} \textbf{298.2}}                                                                    & {\color[HTML]{000000} 12754.4}                                                                                                                                 \\
$u(x, 0)=10$                                    & PPO                                             & 283.2                                                                                                    & 23342.4                                                                                                                                                        \\
$u(x, 0)=10$                                    & SAC                                             & 140.5                                                                                                    & {\color[HTML]{036400} \textbf{9624.1}}                                                                                                                        
\end{tabular}%
}
\caption{Resulting control performance for backstepping, PPO, and SAC on the two PDE examples corresponding to Figure \ref{fig:parabolicExamples}.}
\label{table:parabolicPerformance}
\end{table}

\begin{figure}[H]
    \centering
    \includegraphics{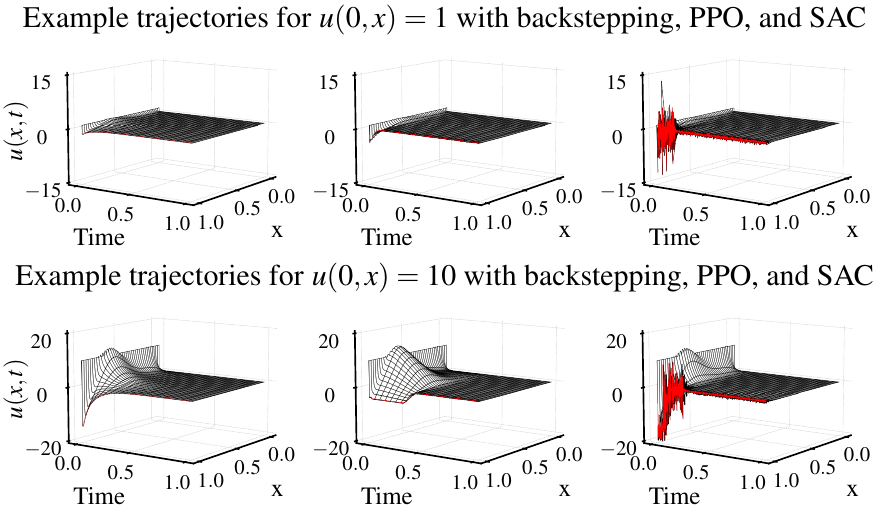}
    \caption{Examples of PDE system stabilization using backstepping, PPO, and SAC (left to right) under two different initial conditions $u(x, 0) = 1, 10$. The PDE has functional recirculation coefficient using the Chebyhsev polynomial defined as $\beta(x)=50\cos(\gamma \cos^{-1}(x))$ with $\gamma=8$. The control signal for each approach is marked in red and in Figure \ref{fig:parabolicControlSignals}.}
    \label{fig:parabolicExamples}
\end{figure}
\vspace{-15pt}
\begin{figure}[H]
    \centering
    \includegraphics{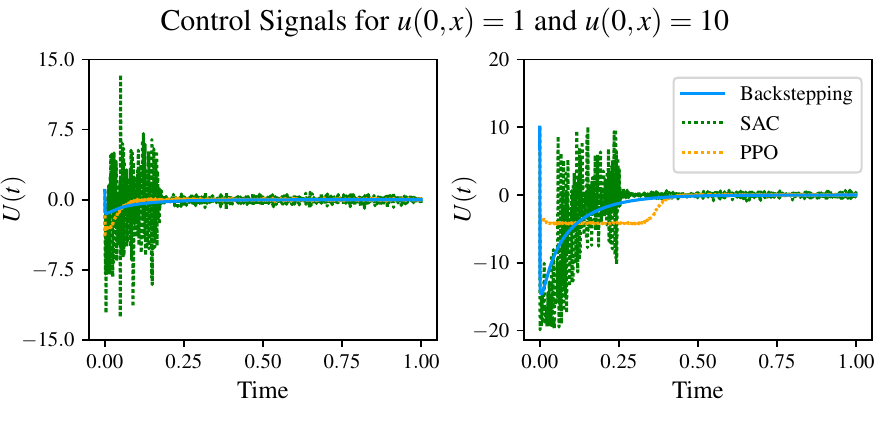}
    \caption{The corresponding control signal given to the PDEs in Figure \ref{fig:parabolicExamples}. The control signals on the left match the PDE with $u(x, 0)=1$ (top-row of Fig. \ref{fig:parabolicExamples}) and the signals on the right correspond to the PDE with $u(x, 0)=10$ (bottom-row of Fig. \ref{fig:parabolicExamples}) respectively.}
    \label{fig:parabolicControlSignals}
\end{figure}

\section{Navier Stokes Equation}
\subsection{Second-Order Finite Difference Scheme}
We consider 2D benchmark Navier Stokes PDE in the form:
\begin{subequations}\label{eq:appendix-ns}
    \begin{align}
    \nabla \cdot \u &= 0 \,, \quad  &&(x, y, t) \in ([0, 1] \times [0, 1]  \times \mathbb{R}_+)\,, \\ 
    \frac{\partial \u}{\partial t} + \u \cdot \nabla \u &= -\frac{1}{\rho} \nabla p + \nu \nabla^2 \u \,, \quad 
    &&(x, y, t) \in ((0, 1) \times (0, 1) \times \mathbb{R}_+)\,,   \\
    \u(x, 0, t) &= U(x,t)\,, \quad &&(x, t) \in ([0, 1] \times \mathbb{R}_+)\,, \\
     \u(x, 1, t) &= \u(1, y, t) = \u(0, y, t) = 0\,, \quad &&(x, y, t) \in ([0, 1] \times [0, 1] \times \mathbb{R}_+)\,,
\end{align}
\end{subequations}
with boundary control input in the top boundary and the velocity at all other boundaries is set to be zero. We incorporate a predictor-corrector~\cite{le1991improvement} scheme for the imcompressive Navier Stokes equations. In the scheme, we denote $\u = (u, v)$ is the velocity vector, and spatial position is $\x = (x,y)$. 

We first present the predictor step that does not consider the pressure, 

\begin{align}
    u^*_{i,j} &= u^n_{i,j} + \Delta t (\nu(\frac{u^n_{i-1,j} - 2 u^n_{i,j} + u^n_{i+1,j}}{(\Delta x)^2} + \frac{u^n_{i,j-1} -2u^n_{i,j}+u^n_{i,j+1}}{(\Delta y)^2})) \nonumber  \\
    &\quad\quad\quad - \Delta t (u^{n}_{i,j}\frac{u^n_{i+1,j}-u^n_{i-1,j}}{2\Delta x} + v^{n}_{i,j}\frac{u^n_{i,j+1} - u^{n}_{i,j-1}}{2 \Delta y})), \\
    v^*_{i,j} &= v^{n}_{i,j} + \Delta t (\nu(\frac{v^{n}_{i-1,j} - 2 v^{n}_{i,j} + v^{n}_{i+1,j}}{(\Delta x)^2} + \frac{v^{n}_{i,j-1} -2v^{n}_{i,j}+v^{n}_{i,j+1}}{(\Delta y)^2}))  \nonumber  \\
  & \quad\quad\quad - \Delta t (u^{n}_{i,j}\frac{v^{n}_{i+1,j}-v^{n}_{i-1,j}}{2\Delta x} + v^{n}_{i,j}\frac{v^{n}_{i,j+1} - v^{n}_{i,j-1}}{2 \Delta y})\,,
\end{align}

where $\Delta t$ denotes the temporal timestep, $\Delta x, \Delta y$ denotes the spatial time step, $n = 0, \ldots Nt$, $i = 0, \ldots, Nx$, $j = 0, \ldots, Ny$ where $Nt,Nx,Ny$ are the total number of timesteps, respectively. $u^{*}_{i,j}, v^{*}_{i,j}$ are the resulted velocity field that does consider pressure. 

To satisfy continuity equation $\nabla \cdot \u = 0$, we get the pressure Poisson equation:
\begin{eqnarray}
    \nabla^2 p = \frac{\partial^2 p}{\partial x^2} +  \frac{\partial^2 p}{\partial y^2} = -\rho (\frac{\partial^2 u}{\partial x^2} + 2 \frac{\partial u}{\partial x}\frac{\partial v}{\partial y} + \frac{\partial^2 v}{\partial y^2})\,.
\end{eqnarray}
Then, we iteratively solve the Poisson equation for pressure:
\begin{align}\label{eq:ns_p}
  p_{i,j}^{iter+1}  & \leftarrow  \frac{1}{2(\Delta x^2 + \Delta y^2)}\left(p^{iter}_{i+1,j}+p^{iter}_{i-1,j})\Delta y^2 
 + (p^{iter}_{i,j+1} + p^{iter}_{i,j-1})\Delta x^2 \right) \nonumber \\
 & \quad + \rho(
  \frac{u^*_{i+1,j} - u^*_{i-1,j}}{2\Delta x} + \frac{v^*_{i+1,j} - v^*_{i-1,j}}{2\Delta y}) \Delta x^2 \Delta y^2\,,
\end{align}
with $iter$ denotes the iteration of the procedure for solving pressure. We denote the pressure solution as $p^*$. Then we can perform the corrector step to compute the velocoty vector at next time step:
\begin{eqnarray}
&  u^{n+1}_{i,j} = u^*_{i,j} - \Delta t \cdot \frac{1}{\rho} \frac{p^*_{i+1,j}-p^*_{i-1,j}}{2\Delta x}\,, \\
&  v^{n+1}_{i,j} = v^*_{i,j} - \Delta t \cdot \frac{1}{\rho} \frac{p^*_{i,j+1} - p^*_{i,j-1}}{\Delta y}\,.
\end{eqnarray}
We apply boundary conditions every time step after the predictor step and corrector step.

\subsection{Control Algorithms Implemented for Navier Stokes PDE}
\subsubsection{Model-Based Optimization}
In optimization algorithms, the optimal control is computed as solution of an optimization problem, where the partial differential equations appear as equality constraints. In the experiment, we consider to track the velocity vector to be the desired trajectory $\u_{ref}(\x,t)$ and to minimize the control cost with a reference control $U_{ref}$. 
The optimization problem (i.e. optimal control) is formulated as follows, 
\begin{subequations}
\begin{alignat}{2}
    \min_{U(t)} \quad & J(U(t), \u),  &&\\
   \text{s.t.} \quad  & \frac{\partial \u}{\partial t} &&= - \u \cdot \nabla \u + -\frac{1}{\rho} \nabla p - \nu \nabla^2 \u, \\
    &\nabla \cdot \u &&= 0, \\ &\u(\x \in \Gamma, t) &&= U(t)
\end{alignat}
\end{subequations}
where $\Gamma$ is the defined controllable boundary. 
The optimization problem can be solved using the augmented Lagrangian:
\begin{eqnarray}    
    L(U, \u, \vec{\lambda}(\x, t), \nu(\x,t)) &=& J(U(t), \u) + \A \mu(\nabla \cdot \u) \nonumber \\ && + \A \vec{\lambda}^T (\frac{\partial \u}{\partial t} + \u \cdot \nabla \u = -\frac{1}{\rho} \nabla p + \nu \nabla^2 \u ).
\end{eqnarray}
A locally optimal solution is characterized by satisfying the first-order Karush-Kuhn-Tucker (KKT) conditions, which necessitate that the derivatives of the Lagrangian function $L$ with respect to variables $U, \u, \vec{\lambda}, \mu$ are all zero. Specifically, taking the derivatives with respect to $\vec{\lambda}$ and $\mu$ yields the Navier-Stokes Equations. Additionally, the solution $\vec{\lambda}$ is required to be divergence-free, a condition that arises from the derivative of $L$ with respect to the pressure $p$. The derivative of $L$ with respect to $\u$ leads to the differential equation for $\vec{\lambda}$:
\begin{equation}\label{eq:nsadjoint}
    \frac{\partial}{\partial t}\vec{\lambda} = -(G + G^T)\u - \nu \nabla^2 \vec{\lambda} - \nabla \mu + (\u - \u_{ref}),
\end{equation}
where $G = \frac{\partial \vec{\lambda}}{\partial \x}$ represents the Jacobian of $\vec{\lambda}$ with respect to $\x$. The differential equation can be solve backwards in time with $\vec{\lambda}(\x, T) = 0$ and homogeneous boundary conditions~\cite{pyta2015optimal}. The derivative with respect to $U$ leads to 
\begin{equation}\label{eq:NSsolution}
    U = U_{ref} - \frac{\nu}{\gamma} \oint_{\Gamma} \frac{\partial \vec{\lambda}_1}{\partial \x_2}.
\end{equation}
Thus the (local) optimal control function $U(t)$ can be computed by 1) Solving Navier Stokes equation to get the velocity vector $\u(x,y,t)$ 2) Solving adjoint equation~\eqref{eq:nsadjoint} for $\vec{\lambda}(x,y,t)$ backwards from time $T$ to $0$ with $\u(x,y,t)$ 3) Solving equation~\eqref{eq:NSsolution} for $U(t)$ forwards with $\vec{\lambda}(x,y,t)$.
\newpage
\subsubsection{Reinforcement learning baselines: proximal policy optimization (PPO) and soft-actor critic (SAC)}
We adopt the exact same, \textbf{untuned}, hyperparameters for PPO and SAC as in Section \ref{sec:hyperbolicRLParams}. 

\subsection{Details on numerical implementation for benchmark 2D Navier Stokes PDE}\label{sec:NSExample}

We adopt the MDP as presented in Table~\ref{table:NSMDP}. For training, we run each model with parameters listed in \ref{table:ppoParameters}, \ref{table:sacParameters} (standard defaults for stable-baselines3, non-optimized) in the environment governed by the MDP in Table \ref{table:NSMDP}. From an implementation perspective, we discretize both the state space using a spatial step of $dx=0.05$ and further discretize each timestep of the MDP by $dt=0.001$ as well. The PDE is internally simulated at a timestep of $dt_{PDE}=dt=0.001$. We use the negative of the cost to be the reward for training. The reference velocity vector $s_{ref}$ is the resulted velocity vector under the boundary control $U(t) = 3 - 5t$, and the reference control $u_{ref}=a_{ref}=2$.

\begin{table}[ht]
\centering
\resizebox{\textwidth}{!}{
\begin{tabular}{l|lll}
\begin{tabular}[c]{@{}l@{}}Markov Decision \\ Process Tuple\end{tabular}                                  & \begin{tabular}[c]{@{}l@{}}Parameter for Navier Stokes\\ PDE Benchmark Example\end{tabular}                                                                                                                                                                                                  &  &  \\ \cline{1-2}
$\mathcal{S}$ - state space                                                                               & $\mathcal{S} \subseteq C([0, 1]\times [0,1])$                                                                                                                                                                                                                                                           &  &  \\
$\mathcal{A}$ - action space                                                                              & $\mathcal{A} \subseteq \{x \in \mathbb{R}: -10 \leq x \leq 10 \} $                                                                                                                        &  &  \\
$\mathcal{O}$ - observation space                                                                         & \begin{tabular}[c]{@{}l@{}}$\mathcal{O} = \mathcal{S}$\\  (can be modified to partially observable)\end{tabular}                                                                                                                                                                          &  &  \\
\begin{tabular}[c]{@{}l@{}}$p_0$ - initial sample\end{tabular}                       & \begin{tabular}[c]{@{}l@{}}$p_0 \subseteq \{f \in C[0, 1]\times[0,1] | f(x,y) = (0,0) \, \forall x \in [0, 1], y \in [0,1]\}$\end{tabular}                                                                                                                                               &  &  \\
$p_f(\cdot | x_t, a_t)$ - state transition                                                                & \begin{tabular}[c]{@{}l@{}}Dynamics described by PDE in \eqref{eq:appendix-ns} \\
with boundary control\, $u(x,0, t)=a_t$ and $v(x,0,t)=0$\end{tabular}                                                                                                                                                         &  &  \\
$T$ - time horizon                                                                                        & $0.2$ seconds                                                                                                                                                                                                                                                                               &  &  \\
$r_a(s, s')$ - reward ($s \in \mathcal{S}$)                                                               & $r_a(s, s') = -\frac{1}{2}* \|s'-s_{ref}\|_{L_2} - \gamma\frac{1}{2}\|a-a_{ref}\|_{L_2}$, $\gamma = 0.1, a_{ref}=2$                                                                                                                                                                                                                                                          &  &  \\
\end{tabular}
}
\caption{Markov Decision Process describing the PDE example in Section \ref{sec:NSExample}}
\label{table:NSMDP}

\end{table}

\end{document}